\documentclass[sigconf]{acmart}
\usepackage{array}
\usepackage{arydshln}
\usepackage{booktabs}
\usepackage{multirow}
\usepackage{xcolor} 
\usepackage{wasysym}
\usepackage{subcaption}

\usepackage{siunitx}
\usepackage{multicol}
\usepackage{boldline}

\usepackage{color,soul}
\DeclareRobustCommand{\hlgreen}[1]{{\sethlcolor{green}\hl{#1}}}

\usepackage{calc}
\usepackage{tikz}

\usepackage{enumitem}

\AtBeginDocument{%
  \providecommand\BibTeX{{%
    \normalfont B\kern-0.5em{\scshape i\kern-0.25em b}\kern-0.8em\TeX}}}

\copyrightyear{2022}
\acmYear{2022}
\setcopyright{rightsretained}
\acmConference[CHI '22]{CHI Conference on Human Factors in Computing Systems}{April 29-May 5, 2022}{New Orleans, LA, USA}
\acmBooktitle{CHI Conference on Human Factors in Computing Systems (CHI '22), April 29-May 5, 2022, New Orleans, LA, USA}
\acmDOI{10.1145/3491102.3501848}
\acmISBN{978-1-4503-9157-3/22/04}

\begin{document}

\title[How Audial Avatar Customization Enhances Visual Avatar Customization]{Audio Matters Too: How Audial Avatar Customization Enhances Visual Avatar Customization}

\author{Dominic Kao}
\email{kaod@purdue.edu}
\affiliation{%
  \institution{Purdue University}
  \country{USA}
}
\author{Rabindra Ratan}
\email{rar@msu.edu}
\affiliation{%
  \institution{Michigan State University}
  \country{USA}
}
\author{Christos Mousas}
\email{cmousas@purdue.edu}
\affiliation{%
  \institution{Purdue University}
  \country{USA}
}
\author{Amogh Joshi}
\email{joshi134@purdue.edu}
\affiliation{%
  \institution{Purdue University}
  \country{USA}
}
\author{Edward F. Melcer}
\email{eddie.melcer@ucsc.edu}
\affiliation{%
  \institution{UC Santa Cruz}
  \country{USA}
}

\renewcommand{\shortauthors}{Dominic Kao et al.}



\begin{abstract}

Avatar customization is known to positively affect crucial outcomes in numerous domains. However, it is unknown whether audial customization can confer the same benefits as visual customization. We conducted a preregistered 2 x 2 (visual choice vs. visual assignment x audial choice vs. audial assignment) study in a Java programming game. Participants with visual choice experienced higher avatar identification and autonomy. Participants with audial choice experienced higher avatar identification and autonomy, but only within the group of participants who had visual choice available. Visual choice led to an increase in time spent, and indirectly led to  increases in intrinsic motivation, immersion, time spent, future play motivation, and likelihood of game recommendation. Audial choice moderated the majority of these effects. Our results suggest that audial customization plays an important enhancing role vis-\`a-vis visual customization. However, audial customization appears to have a weaker effect compared to visual customization. We discuss the implications for avatar customization more generally across digital applications.

\end{abstract}

\maketitle

\section{Introduction}

Avatars are ubiquitous across digital applications. Using avatars as representations of ourselves, we socialize, play, and work. Increasingly, researchers have become interested in \textit{avatar customization} \cite{mcarthurUX2017}. Avatar customization, or the ability to modify one's avatar, increases outcomes including intrinsic motivation \cite{birkFostering2016}, helping behavior \cite{dolgov2014effects}, user retention \cite{birkCombating2018}, learning \cite{ngExamining2013}, flow \cite{liao2019avatar}, and of especial importance, avatar identification \cite{turkay2015effects}. Avatar identification, or \textit{the temporary alteration in self-perception of the player induced by the mental association with their game character} \cite{vanlooyPlayer2012}, leads to increased motivation \cite{birkFostering2016, turkayEffects2014, birkCombating2018}, creative thinking \cite{derooijCreative2017, buisineUsing2016, gueganAvatarmediated2016}, enjoyment \cite{trepteAvatar2010, ngExamining2013, liPlayer2016}, performance \cite{kaoEffects2018}, player experience \cite{kao2019effects}, flow \cite{soutterRelationship2016}, and trust in others \cite{kimBecame2012}. Despite the  large corpora of literature on avatar customization, studies have focused almost exclusively on \textit{visual} aspects of the avatar. Limited adoption of \textit{audial} aspects in avatar customization is potentially because avatar audio is perceived as non-critical and has substantial overhead  (e.g., multiple voice actors, region localization) \cite{5b7cb5a5ef1a40df91b7db27713459e5}. Recent advances in artificial intelligence (e.g., neural networks) have vastly improved text-to-speech engines and voice cloning software, however, and these programs are able to produce artificial voices nearly indistinguishable from real ones. Voice cloning software was used in a study in which participants played a game with avatars that had either a similar voice or a dissimilar voice (as compared to the player) \cite{kao2021chiplayavatarvoice}. Results showed that participants in the similar voice condition had increased  performance, time spent, similarity identification, competence, relatedness, and immersion. Prior research adds further support that the importance of avatar audio may be underappreciated. Audio in games is linked to increased physiological responses \cite{Hebert2005}, emotional realism \cite{Berndt2008,Ekman2008}, performance \cite{Johanson2016}, and immersion \cite{Ekman2013,Larsson2010,Nacke2011,sanders2010time,keehl2019radical}. A meta-analysis of 83 studies in virtual environments found that adding audio has a small- to medium-sized effect on presence \cite{Cummings2016}. Given prior work demonstrating the importance of avatar customization and audio separately, allowing players to \textit{audially} customize their avatars may have beneficial effects.

Customizing one's avatar is often viewed as inherently enjoyable \cite{kafaiYour2010}. This customization is now part of a lucrative ``skin'' market in online games \cite{jarrettGaming2021}. Game skins can be used to customize an avatar's appearance, and research estimates the skin market to be worth \$40 billion (USD) per year \cite{skins40billion}. While a few ventures have begun to explore customization of the player's voice, these efforts have been limited to external tools (e.g., voice-changing software \cite{Voicemod2020,modulatevoicewear}). A small number of games \textit{do} offer the option of customizing avatar audio. Final Fantasy XIV \cite{SquareEnix2013}, Saints Row IV \cite{Volition2013}, and Monster Hunter: World \cite{Capcom2018} allow the user to choose between different sets of voices. Black Desert Online \cite{PearlAbyss2015}, Red Dead Redemption 2 \cite{RockstarGames2018}, and The Sims 4 \cite{ElectronicArts2014} allow the user to customize pitch. More generally, avatar customization interfaces are understood to vary greatly between games with regards to both quantity and quality of customization options \cite{mcarthurAvatar2015, mcarthurUX2017}. For the purposes of the present study, we created four character models and four character voices. We then created four character customization interfaces that varied (1) whether the character model was chosen or randomly assigned and (2) whether the character voice was chosen or randomly assigned. These customization interfaces were explicitly designed to test whether audial customization would have any effect on outcomes vis-\`a-vis visual customization.

We conducted an online study on Amazon's Mechanical Turk (MTurk) in which participants were randomly assigned to one of the four character customization interfaces. Participants then played a Java programming game for 10 minutes. After 10 minutes had passed, an in-game survey collected measures of avatar identification, autonomy, intrinsic motivation, immersion, motivated behavior, motivation for future play, and likelihood of game recommendation.\footnote{Study hypotheses, analyses, experiment design, data collection, sample size, and measures were all preregistered. \\ Preregistration: \url{https://osf.io/dbvkp/}.\\ Raw Data: \url{https://osf.io/mnpsd/}.} After completing the survey, participants could quit or continue playing for as long as they liked, reflecting  motivated behavior.

Our results show that visual customization leads to higher avatar identification and autonomy. Audial customization leads to higher avatar identification and autonomy, but only within the grouping of participants in which visual customization was available. In the grouping of participants without visual customization, audial customization had no effect on avatar identification or autonomy. Visual customization leads to higher time spent playing, and indirectly (through the mediators of avatar identification and autonomy), it leads to higher intrinsic motivation, immersion, time spent playing, motivation for future play, and likelihood of game recommendation. Audial customization moderated the direct effect of visual customization on time spent playing, as well as the indirect effects of visual customization on intrinsic motivation, immersion, motivation for future play, and likelihood of game recommendation. The moderation effect was such that the effect was non-significant when audial customization was unavailable but significant when audial customization was available. Our results show that audial customization, although having an overall weaker effect than visual customization, can strengthen existing effects of visual customization on outcomes. This suggests that avatar customization systems in games can be improved by adding audial customization options. Moreover, our study provides motivation to extend this research to other domains as potential beneficiaries of audial avatar customization (e.g., virtual reality, digital learning, health applications). In the highly understudied area of avatar audio, we contribute baseline results in a large-scale preregistered study that can spur further work in this domain.

\section{Related Work}

\subsection{Avatar Customization}

Avatar customization is the process of changing aspects of a video game character. Players customize their avatars' physical (e.g., body shape), demographic (e.g., age, race, gender), and transient (e.g., clothes, ornaments) aspects. The avatar customization process can also include choosing roles (e.g., playing as a warrior, archer, mage, or a healer), attributes (e.g., luck, intelligence), and group membership (e.g., playing as horde or alliance) \cite{turkayFree2010,WoW}. Customizing one's avatar can lead to direct and indirect effects on gameplay \cite{turkayFree2010,isbister2006better}. For example, choosing a role of a warrior affords different game mechanics and play strategies (i.e., favoring close combat) compared to playing as an archer. Similarly, customizing skill attributes can also affect gameplay---e.g., favoring increased charisma gives lower prices on game items in \textit{Fallout 4} \cite{Fallout4}. Customizing avatars' physical appearance or the name of the avatar, on the other hand, usually does not affect gameplay (directly) but can have a psychological effect on the players \cite{birkFostering2016, schmierbachFeeling2012, limBeing2009}. To understand these psychological effects, many studies have used off-the-shelf games (e.g., Massively Multiplayer Online Games, or MMOs) that offer a comprehensive avatar customization process, such as changing physical, demographic, and transient aspects; as well as choosing roles, group membership, and attributes. Lim and Reeves used a popular MMORPG (\textit{World of Warcraft, or WoW} \cite{WoW}) where participants were randomly assigned to play the game with avatar customization or to play with a premade avatar \cite{limBeing2009}. The study found that players who customized their avatar experienced greater physiological arousal \cite{limBeing2009}. Similarly, players reported greater physiological arousal and subjective feelings of presence when playing advergames that offered avatar customization options, suggesting greater game enjoyment \cite{baileyHow2009}. It has also been shown that players remember more game features---such as spatial features of landmarks and characteristics of NPCs---when playing with customized avatars \cite{ngExamining2013}. Teng \cite{teng2021can} examined how customizing avatars' transient aspects in MMORPGs impact identification and loyalty with the game. The study found that customizing these items (e.g., clothes, shoes, etc.) positively impacted identification with the avatar, which subsequently increased gamer loyalty. Other studies have also explored how customizing non-human objects (e.g., race cars) influences player experience \cite{schmierbachFeeling2012,RatanCompanionsVehicles2018}. One study used the game \textit{Need for Speed: ProStreet} \cite{NSFProstreet} to understand if customizing a racecar affects players' enjoyment of the game \cite{schmierbachFeeling2012}. Players customized their cars' visual appearance, such as changing the car's shape, aftermarket components (spoilers, rims), color, and skins. Players who customized their cars experienced greater identification, leading to higher game enjoyment, than those who played with pre-made customized cars. One key limitation of these studies is the time duration of their investigation. Many studies have only investigated the effect of avatar customization on short playing time (\textasciitilde 1 hour) \cite{turkayEffects2014}. MMOs are long-term games, with players' gameplay experience and expertise evolving with time. Previous studies have found that players playing these games spend approximately 10 hours playing each week \cite{ducheneautAlone2006}. Turkay and Kinzer investigated how players' identification and empathy towards their avatar evolved over ten hours of playing \textit{Lord of the Rings Online} (\textit{LotRO} \cite{LotRO}) \cite{turkayEffects2014}. The study found that players who customized their avatars had a stronger identification and expressed greater empathy towards them than those who played the game with premade avatars. 

Studies have also used bespoke games to understand the effects of avatar customization \cite{birkFostering2016, linHow2017, koulourisMe2020}. Birk, Atkins, Bowey, and Mandryk \cite{birkFostering2016} investigated if players who customized their avatars experienced greater intrinsic motivation compared to those who used premade avatars. The researchers leveraged \textit{Unity Multipurpose Avatar} \cite{UMA} to develop a character creator which allowed players to customize their game characters' appearance (e.g., skin tone, clothing), personality (e.g., extraversion), and attributes (e.g., intelligence, stamina, willpower). Players who customized their game character experienced greater identification with their avatars, which led to greater autonomy, immersion, invested effort, enjoyment, positive affect, and time spent playing in an infinite runner \cite{birkFostering2016}. In a subsequent paper, Birk and Mandryk investigated the effect of avatar customization on attrition and sustained engagement while playing a mental health game over three weeks \cite{birkCombating2018}. The study found a reduced attrition rate for the players who customized their avatar compared to those who played with a generic avatar \cite{birkCombating2018}. In another study, playing an exergame with autonomy-supportive features (which included customizing an avatar) led to increased effort, autonomy, motivation to play the game again, and greater likelihood to recommend the game to peers compared to participants who played the game without autonomy-supportive features \cite{Peng2012}. Similarly, in a virtual reality exergame, players customized their avatars using an off-the-shelf software tool (\textit{Autodesk Character Creator} \cite{autodesk}) to create an avatar similar to themselves. Players could customize their avatars (e.g., skin tone, hair and eye color, clothes, shoes). The study found that players who competed against their customized self-similar avatars performed significantly better compared to the players who competed with generic avatars \cite{koulourisMe2020}. The effect of customization has also been observed in learning environments. Students engaged with a computational learning game (over seven sessions lasting an hour each) with a customized avatar of their choosing \cite{linHow2017}. Customization options included skin tone, hairstyle, and eye-color options. The study found that players who customized their avatars remembered and understood greater computational concepts than those who played the game with a premade avatar. Kao and Harrell \cite{kaoEffects2018} investigated how avatar identification influenced players in a computational learning environment (\textit{MazeStar} \cite{Kao2017b}). Players customized their avatars using a freely available Mii creator. The study found that avatar identification promoted outcomes including player experience, intrinsic motivation, and time spent playing \cite{kaoEffects2018}.

These studies suggest that avatar customization affects player experience in a wide variety of settings (e.g., games for entertainment or learning), virtual environments (e.g., desktop, VR) and timespans (both one-off play sessions and longitudinal) \cite{birkFostering2016,schmierbachFeeling2012,limBeing2009,koulourisMe2020, turkayEffects2014, kao2021chiplayavatarvoice}. More importantly, a subset of these studies highlight that avatar customization generates attachment and identification with their game character \cite{birkCombating2018, turkayEffects2014, schmierbachFeeling2012,teng2021can, kaoEffects2018}, which consequently affects a wide range of variables: intrinsic motivation \cite{Peng2012, birkFostering2016}, autonomy \cite{birkFostering2016, birkCombating2018, kao2021chiplayavatarvoice}, empathy \cite{turkayEffects2014}, performance \cite{koulourisMe2020, birkCombating2018, kao2021chiplayavatarvoice}, game enjoyment \cite{trepteAvatar2010}, loyalty \cite{teng2021can} and player experience \cite{birkFostering2016,schmierbachFeeling2012,limBeing2009, turkayEffects2014, kao2021chiplayavatarvoice}.  

\subsubsection{Avatar Identification}
Identification is a mechanism wherein media experiences---such as reading a story or watching a movie---are interpreted and experienced by audiences as if ``the events were happening to them'' \cite{cohenDefining2001}. The mechanism of identification differs in interactive and non-interactive media experiences. In a typical media experience (e.g., movie or a late-night talk show), the relationship between the audience and media-character is often categorized as a \emph{self} versus \emph{other} (often referred to as a dyadic relationship) \cite{downsPolythetic2019, christophVideo2009}. Within games, the distance between the self and the other is said to be diminished due to games' affording direct control over the game character and their interactions in the virtual world \cite{klimmt2003dimensions,Hefner2007}. Players control, customize, and interact with their game character and the game world using an avatar. Consequently, the player-avatar relationship is often said to be ``a merging of [the player] and the game protagonist'' \cite{christophVideo2009}.

Avatar identification is thought to be a shift in self-perception \cite{klimmt2010identification}. Players can temporarily adopt salient characteristics of the avatar \cite{vanlooyPlayer2012} or channel their expectations into the avatar creation, thereby facilitating avatar identification \cite{turkayEffects2014}. Many factors influence the nature of identification that can take place with the avatar. Flanagan \cite{flanaganMobile1999} asserts that player identification with a game character is complicated by the various roles embodied by the player (such as being a subject, spectator, participant, etc.) during gameplay. Murphy \cite{murphy2004live} elaborates on how players' abilities, player characters' abilities, game events, and other players influence the player's sense of agency in virtual environments. While many authors agree that identification takes place between a player and the game character, the nature of identification remains understudied \cite{turkayEffects2014}. 

One avenue of understanding identification is through understanding the avatar customization process. When  players customize their avatar, they cycle through many ``possible selves'' \cite{markus1986possible} as they experiment and adopt the game characters' attributes for themselves. In two separate studies by different researchers, there are a few common trends regarding players' avatar creation and customization experiences
 \cite{kafaiYour2010,ducheneautBody2009}. In one of the studies, researchers investigated reasons for avatar customization and creation in three virtual worlds: \textit{World of Warcraft} \cite{WoW}, \textit{Second Life} \cite{SecondLife} and \textit{Maple Story} \cite{MapleStory}. Researchers found that players in these virtual worlds created and customized their avatars for various reasons, including to project an ideal self, follow a trend, or stand out from others \cite{ducheneautBody2009}. Another study examined the avatar creation and customization process for players in \textit{Whyville} \cite{kafaiYour2010}. Players customized their avatars for aesthetic reasons, to follow a popular trend, and to express themselves (e.g., show some aspect of their authentic selves). Moreover, they also found that players customize their avatars with a functional intention, such as to experiment with gender or to play different roles \cite{kafaiYour2010}. 

These findings have led researchers to consider avatar identification as a multi-faceted construct \cite{downsPolythetic2019}, which has been operationalized into three distinct dimensions: \emph{similarity identification}, \emph{wishful identification}, and \emph{embodied identification} \cite{vanlooyPlayer2012}. Similarity identification refers to players identifying with an avatar that looks like them \cite{downsPolythetic2019}. Avatars that look similar to players can facilitate feelings of familiarity and stronger empathetic experience \cite{vanlooyPlayer2012}. Research shows that similarity identification can play an important role in the player's motivations for playing \cite{vanlooyPlayer2012}, learning outcomes \cite{kao2021chiplayavatarvoice}, player experience \cite{kao2019effects}, and behaviors \cite{liPlayer2016,koulourisMe2020}. Players can also identify with their game characters and see them as role models for future action or identity development \cite{vanlooyPlayer2012}. Players desiring to align their personal attitudes, aesthetics, and attributes with those of their game character is referred to as wishful identification \cite{vanlooyPlayer2012, downsPolythetic2019}. For example, previous research has documented that older players often create avatars younger than themselves \cite{ducheneautBody2009}. Lastly, players also identify with their avatars when manipulating avatars' bodies as their own. Perceiving to be present in a virtual environment through one's avatar, or so-called ``body container'' \cite{vanlooyPlayer2012, downsPolythetic2019}, heightens embodied identification \cite{turkayEffects2014}.

The process of avatar customization is often a precursor for generating greater avatar identification. For example, players wanting to create an avatar that has similar attributes (e.g., physical appearance, hair style, hair color) may generate greater similarity identification \cite{turkayEffects2014}. On the other hand, players customizing their avatars according to their ideal self may increase their wishful identification \cite{turkayEffects2014}. Players typically interact with a user interface that allows players to fluidly cycle through choices to allow players to constitute their desired digital body. As such, the design and options presented to the players can play a crucial role in helping (or hindering) players to create their desired avatar \cite{mcarthurEveryone2014, mcarthurChallenging2018}. 

\subsubsection{Avatar Customization Interface}

The interface that the players use to create and customize their avatars---sometimes referred to as a character customization interface (CCI) \cite{mcarthurAvatar2015}---represents a ``space of liminality" \cite{waggoner2007passage} where players spend a significant amount of time intentionally creating their desired avatar \cite{ducheneautBody2009, mcarthurAvatar2015}. McArthur states that these interfaces generate action possibilities for avatar creation and customization \cite{mcarthurAvatar2015}. Players cycle through many possible customization options to create their desired avatar. Avatar customization interfaces are not only important in terms of usability, but also in how they communicate cultural ideologies \cite{mcarthurAvatar2015, mcarthurChallenging2018}.

For instance, the design of ``default'' options in avatar customization interfaces and the order (hierarchy) of body customization options oftentimes implicitly reinforces existing hegemonic structures in society \cite{mcarthurAvatar2015, nakamuraCybertypes2013}. Avatar customization interfaces are known to constrain user choices, in part due to their oftentimes exclusionary design \cite{mcarthurEveryone2014}. Previous research has found a limited number of options for players belonging to diverse ethnic groups and gender, suggesting that customization favors the creation of light-skinned male avatars \cite{consalvoIt2003, paceAre2009, mcarthurAvatar2015}. While our focus in the present study is on understanding if audial avatar customization can confer similar benefits to visual avatar customization, the exclusionary potential of audial avatar customization options should be studied closely in future research.

Research has emphasized the role played by other aspects including game world aesthetics, co-situated players, social context, and avatars of other characters in influencing the avatar customization process \cite{mcarthurChallenging2018, kafaiYour2010}. Kafai found that new players felt out of place with their generic avatars when interacting with avatars with detailed customization \cite{kafaiYour2010}. Players also reported customizing their avatars to avoid being bullied in online settings by other players \cite{kafaiYour2010}. Players customize their avatars differently depending on the context of the virtual environment, such as changing clothes and accessories when the social context switched between ``game'' and ``job'' \cite{tribertiChanging2017}. Players also adhere to group norms while creating and customizing their character \cite{jaimebanksBard2017}. User characteristics, such as age, gender and self-esteem, play a role in the avatar creation and customization process. Individuals with higher self (and body) esteem represent their avatars with a greater number of body details and emphasis on sexual characteristics that identified their gender \cite{villaniExploration2016}. Adolescent boys customized their avatars to create a more stereotypical masculine body compared to girls who focused on customizing transient aspects of the avatar, such as clothing and accessories \cite{villaniExploration2016}. 

Although the process of avatar customization has been extensively investigated, research has largely ignored the effect of voice options on avatar creation and customization. Contemporary games seldom offer voice customization options; however, there do exist some examples. Some games offer a ``voice template'' that can be chosen during avatar customization, such as in  \textit{Black Desert Online} \cite{PearlAbyss2015}. \textit{Sims 4} \cite{ElectronicArts2014} allows characters' voices to be customized according to three voice qualities: ``sweet,'' ``melodic,'' and ``lilted'' for women, and ``clear,'' ``warm,'' and ``brash'' for men. Other games allow players to customize a given voice by directly changing specific aspects of the voice, such as pitch. The games \textit{Saints Row IV} \cite{Volition2013} and \textit{Cyberpunk 2077} \cite{Cyberpunk2077} offer the ability to modify pitch. This project investigates the effect of providing audial avatar customization options on a variety of player outcomes.

\subsection{Audio in Games}

Game audio performs many functions, such as emphasizing visuals \cite{neuholdRole}, contextualizing a place \cite{Ekman2013}, highlighting emotions and thoughts of the game-character \cite{neuholdRole}, and immersing the player in the game world \cite{rogersExploring2017}. To understand the design of audio in games, researchers have defined audio typologies. One typology classifies sound based on the source \cite{berndtDiegetic2011}. Sound is referred to as ``diegetic'' if it originates from the game world (e.g., game sound \cite{grimshaw2008sound, nackeMore2010}), and sounds that have origins different than the game world (e.g., interface sounds) are called ``non-diegetic'' \cite{grimshaw2008sound, nackeMore2010}. Liljedahl \cite{liljedahlSound2011} classifies sounds into three categories: speech and dialogue, sound effects (e.g., ambient noise, avatar sounds, object and ornamental sounds), and music. 

Research shows that players appreciate the inclusion of audio elements in the game. Klimmt et al. \cite{klimmtEffects2019} investigated the role of background music on gameplay experiences of players. Players experienced greater enjoyment while playing a game (\textit{Assassins Creed: Black Flag} \cite{AC4}) with background music included. Background music can also affect performance in a game---participants who played a role-playing adventure game (\textit{The Legend of Zelda: Twilight Princess} \cite{ZeldaTP}) performed better with background music present \cite{siu-lantanEffects2010}. Some games incorporate background music that changes according to events in games. An adaptive soundtrack has also been shown to improve player experience. Researchers designed a game with a soundtrack that increased in tension depending on the chance of success or failure of players in the game \cite{plutMusic2019}. Participants who played the game with an adaptive soundtrack experienced greater tension, suggesting a more engaging experience. Players playing a first-person shooter game reported higher game experience (immersion, flow, positive affect) with the presence of sound effects (e.g., ornamental and character sounds) \cite{nackeMore2010}. Audio may also influence motivated behaviors such as time played \cite{kao2020effects} and actions performed \cite{kao2019infinite}. Lack of thematic fit between audio and visuals (also known as game atmosphere) can affect player experience. In a study, players played a survival horror game (\textit{Bloodborne} \cite{Bloodborne}) either with background and voiceover audio relevant for the game (built-in game audio) or with experimenter-induced music and voiceovers \cite{giovanniribeiroGame2020}. Players experienced a lower degree of perceived game atmosphere when the audial elements did not fit the game's visual elements. 

Avatar sounds are sounds related to the avatar activity, such as breathing and footstep sounds \cite{liljedahlSound2011}. These sounds help immerse the player into the game world  \cite{grimshawSound2007}, provide feedback for avatar movement \cite{Ekman2013}, and play a crucial role in localizing the player in audio games \cite{fribergAudio2004, garciaDesign2013}. For example, Adkins et al. \cite{adkinsLost2020} developed an audio game wherein the players selected an animal as a game character---a cow, dog, cat, and frog---to navigate through a maze. The four animals also had representative animal sounds that provided essential user feedback for nearby obstacles and intersections. Providing sound cues for the movement of an avatar helps the virtual world conform to the players' expectations \cite{fribergAudio2004} and induce immersion into the game world \cite{grimshawSound2007}. 

\subsubsection{Avatar Voice} 

Avatar voice includes linguistic (e.g., dialogue and voiceovers) and non-linguistic vocalizations such as emotes (e.g., effort grunts, screams, sighs) \cite{holmesDefining2021}. Avatar voice can be used to control actions of the game character \cite{allisonDesign2018}, converse with NPCs \cite{domsch2017dialogue}, and converse with other players in the game world \cite{wadley2015voice}. While conversation with NPCs is usually supported through prerecorded dialogues \cite{holmesDefining2021}, games also facilitate avatar control and player-to-player communication through voice interaction \cite{carterPlayer2015}. 

Voice dialogue in games supports storytelling, the development of a rich and believable world, and setting emotional tone \cite{holmesDefining2021}. As players explore and interact with a novel game world, conversing with NPCs can reveal important information regarding historical events and new quests that can ultimately help in the narrative progression. A common feature in many open-world games is the presence of a social space (e.g., local tavern) containing music and ambient sounds that are concurrent and continuous \cite{smuckerGaming2018}. The social space also contains jumbled, indistinct conversations (Walla) among social actors (NPCs) \cite{holmesDefining2021}. Therefore, a sonic environment comprising of music, sounds, and voices helps in several ways: creating a game-feel, setting the mood, and making the game world believable \cite{collins2008game, holmesDefining2021}. Game characters also use emotionally-laden dialogues to engender emotions in a player that can forge a deeper connection between the game character and the player \cite{stockburger2010play}. For instance, an urgent request for help can arouse the player to take action.  

Voice interaction focuses on using players' voices as input in the game \cite{carterPlayer2015, allisonDesign2018}. Beyond using voice interaction to converse with other players \cite{wadley2015voice}, recent advances in software and hardware technology \cite{allisonWord2020} have made it possible to use voice interaction to control avatar actions and in-game events \cite{allisonDesign2018}. Two popular approaches exist here: ``voice-as-sound'' \cite{igarashi2001voice,hamalainenMusical2004} and ``voice-as-speech'' \cite{allisonDesign2018, carterPlayer2015}. Voice-as-sound uses players' voice characteristics such as pitch and tone \cite{igarashi2001voice,hamalainenMusical2004}. H{\"a}m{\"a}l{\"a}inen et al. \cite{hamalainenMusical2004} describes the design of two games that used the voice-as-sound approach. The players navigate a hedgehog through a maze in the first game by singing at the correct pitch. The authors also developed a turn-based ping-pong game where the players had to navigate their paddle at appropriate positions using the correct pitch. Voice-as-speech uses speech recognition technology to interpret players' commands in games \cite{allisonWord2020, hamalainenMusical2004, allisonDesign2018}. Players can use their voice to navigate menus \cite{ carterPlayer2015}, engage in unscripted conversations with a virtual pet (a fish in the game \textit{Seaman} \cite{Seaman}) \cite{allisonDesign2018}, and cast spells using voice commands in \textit{Skyrim} \cite{Skyrim, allisonWord2020}.  

Carter suggests that voice interaction can facilitate a deeper connection with the players' game characters \cite{carterPlayer2015}. The voice of an avatar is a part of game characters' identity, and providing a way to use players' voices for avatar actions can lead to a merging of identity (player-avatar convergence). Embodied identification, that is, the degree of control over the game characters' movement and action, can imbue players with a greater sense of agency and identification \cite{turkayEffects2014}. Players playing \textit{Tomb Raider} \cite{TombRaider} can use voice commands to initiate player actions, such as attack and defend \cite{carterPlayer2015}, while simultaneously performing (other) actions with the game controller. In this sense, voice interaction may facilitate embodied identification by affording greater control over game characters' actions. Voice interaction may also facilitate wishful identification by affording associations between players' voice and the game characters' voice. \textit{Splinter Cell Blacklist} \cite{Blacklist} allows users to distract enemy NPCs by using a specific speech phrase (``Hey, you!''), which is repeated by the voice of the game character in the virtual world. Players in \textit{FIFA 14} \cite{FIFA14} embody the role of manager and perform actions such as selecting players for the tournament and giving advice on the field. Players can voice specific commands that change the behavior of their chosen team to adopt a defensive or attacking mindset \cite{carterPlayer2015}---a typical action that coaches and managers perform. Lastly, voice interactions can facilitate similarity identification by allowing users to interact with the game characters using their voices. For example, avatar representation in karaoke games is almost entirely through the voice of the player \cite{carterPlayer2015}. 

\subsubsection{Avatar Voice and Learning Environments}

Studies have investigated how engagement and learning outcomes are influenced by voice characteristics of the instructional agents \cite{lee2007children, mayer2003social}. Learners rate voices more likeable when voice characteristics of instructional agents are similar to themselves in perceived gender \cite{lee2000can} or personality \cite{nassDoes2000,nassDoes2001}. Research also documents persistent stereotypes in the design of instructional agents' voices. Deutschmann \cite{deutschmannGenderbending2011} evaluated how students perceive a male and female avatar delivering a lecture. Students perceived the male avatar as more knowledgeable, and the female character as more likable. Along a similar line, authors designed three avatars---the instructor's face, male-anime, and female-anime---to understand how students perceive and perform in an online course. Students showed higher likeability for the female-anime avatar but performed higher when instructors' own face delivered lectures. Although these studies show that the voice of an avatar plays a role in students' perception and performance, a general limitation is the poor quality of voice morphing in these studies \cite{deutschmannGenderbending2011, hsiehEvaluation2021}. 

More recently, research has also sought to understand how an avatar's voice can affect self-presentation in digital environments. Zhang et al. characterized users' voice customization preferences on social media websites \cite{zhangSocial2021}. The study highlighted gender, personality, age, accent, pitch, and emotions as key factors that users wanted to customize to represent their avatar in digital spaces \cite{zhangSocial2021}. The study also highlighted the need to provide customization options to modulate pitch and voice depending on the context---e.g., sounding serious and formal for professional websites such as LinkedIn. A common trend in studies leveraging personalized avatar voice in virtual environments is the beneficial effects of using a self-similar avatar voice \cite{kao2021chiplayavatarvoice, aymerich-franchEffects2012}. In a public speaking experiment, participants stood in front of a virtual classroom to give a speech \cite{aymerich-franchEffects2012}. Participants either used their own voice to give the speech or had another participant's speech played back. Participants who used their own voice showed significantly higher social presence \cite{aymerich-franchEffects2012}. Kao, Ratan, Mousas, and Magana leveraged recent advances in voice cloning and found that learners using a more self-similar voice (as opposed to a self-dissimilar voice) in a game-based learning environment had higher performance, time spent, similarity identification, competence, relatedness, and immersion. Additionally, they found that similarity identification was a significant mediator between voice similarity and all measured outcomes \cite{kao2021chiplayavatarvoice}.

While research provides strong support for avatar voice influencing avatar identification, no study (to the best of our knowledge) has investigated the effects of providing avatar audial customization options. We present a study that provides audial (voice) avatar customization options alongside visual avatar customization options in a Java programming game. Our goal is to understand how providing audial avatar customization options affect measured outcomes.

\subsection{Hypotheses}

We had seven overarching hypotheses (each broken down into three sub-hypotheses) in this study. All hypotheses and research questions were part of the study preregistration.\footnote{\url{https://osf.io/dbvkp/}.} Because prior work has shown that avatar customization leads to an increase in avatar identification (similarity identification, embodied identification, and wishful identification)  \cite{birkFostering2016,birkCombating2018,dechant2021chiplayavatar}, we hypothesized that visual customization would lead to an increase in avatar identification. Research has shown that game audio is important to player experience (PX) \cite{Ekman2008,Nacke2011,Ekman2013} and that avatar audio can influence avatar identification \cite{kao2021chiplayavatarvoice}. Therefore, we hypothesized that audial customization would lead to an increase in avatar identification. Additionally, we hypothesized a lack of an interaction effect between visual and audial customization because existing work gives us no reason to believe their effects would depend on one another.

\setitemize{noitemsep,topsep=3pt,parsep=0pt,partopsep=0pt}

\begin{itemize}
    \item[\textbf{H1.1}:] Visual customization will lead to higher avatar identification.
    \item[\textbf{H1.2}:] Audial customization will lead to higher avatar identification.
    \item[\textbf{H1.3}:] No interaction effect between visual and audial customization for avatar identification.
\end{itemize}

Prior studies have shown that character customization leads to greater autonomy \cite{Peng2012,birkFostering2016}. Therefore, we hypothesized that visual customization would lead to greater autonomy. Similar to H1.2, we hypothesized that audial customization will play a similar role to visual customization and will also increase autonomy. We again hypothesized a lack of an interaction effect for the same reason as H1.3.

\begin{itemize}
    \item[\textbf{H2.1}:] Visual customization will lead to higher autonomy.
    \item[\textbf{H2.2}:] Audial customization will lead to higher autonomy.
    \item[\textbf{H2.3}:] No interaction effect between visual and audial customization for autonomy.
\end{itemize}

Prior work has shown that avatar customization is linked to intrinsic motivation \cite{birkFostering2016}, immersion \cite{birkFostering2016}, time spent playing \cite{birkFostering2016}, motivation for future play \cite{Peng2012}, and likelihood of game recommendation \cite{Peng2012}. Furthermore, avatar identification and autonomy are increased through avatar customization (e.g., \cite{birkFostering2016,Peng2012}), and also affect intrinsic motivation, immersion, time spent playing, motivation for future play, and likelihood of game recommendation  \cite{birkFostering2016,kaoEffects2018,Peng2012,Przybylski2010,Ryan2006}. Therefore, we hypothesized a model in which visual customization directly, and indirectly through avatar identification and autonomy, influences intrinsic motivation, immersion, time spent playing, motivation for future play, and likelihood of game recommendation. Lastly, given the lack of prior work on audial customization, we posed as research questions (without any formal hypotheses) whether audial customization moderated any of these effects.

\begin{itemize}
    \item[\textbf{H3.1}:] Visual customization will lead to higher intrinsic motivation.
    \item[\textbf{H3.2}:] Avatar identification will mediate H3.1.
    \item[\textbf{H3.3}:] Autonomy will mediate H3.1.
\end{itemize}

\begin{itemize}
    \item[\textbf{H4.1}:] Visual customization will lead to higher immersion.
    \item[\textbf{H4.2}:] Avatar identification will mediate H4.1.
    \item[\textbf{H4.3}:] Autonomy will mediate H4.1.
\end{itemize}

\begin{itemize}
    \item[\textbf{H5.1}:] Visual customization will lead to higher time spent playing.
    \item[\textbf{H5.2}:] Avatar identification will mediate H5.1.
    \item[\textbf{H5.3}:] Autonomy will mediate H5.1.
\end{itemize}

\begin{itemize}
    \item[\textbf{H6.1}:] Visual customization will lead to higher motivation for future play.
    \item[\textbf{H6.2}:] Avatar identification will mediate H6.1.
    \item[\textbf{H6.3}:] Autonomy will mediate H6.1.
\end{itemize}

\begin{itemize}
    \item[\textbf{H7.1}:] Visual customization will lead to higher likelihood of game recommendation.
    \item[\textbf{H7.2}:] Avatar identification will mediate H7.1.
    \item[\textbf{H7.3}:] Autonomy will mediate H7.1.
\end{itemize}

\noindent \textbf{Research Question}: Does audial customization moderate H3--H7?

\begin{figure*}
\centering
\begin{subfigure}{.49\textwidth}
  \centering
  \includegraphics[width=1\linewidth]{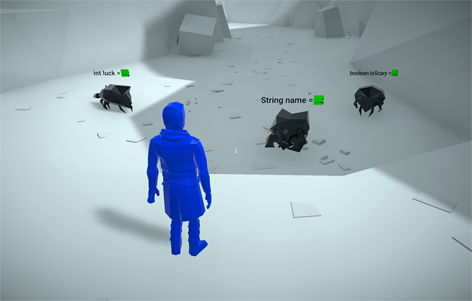}
\end{subfigure}%
\hspace*{\fill}
\begin{subfigure}{.49\textwidth}
  \centering
  \includegraphics[width=1\linewidth]{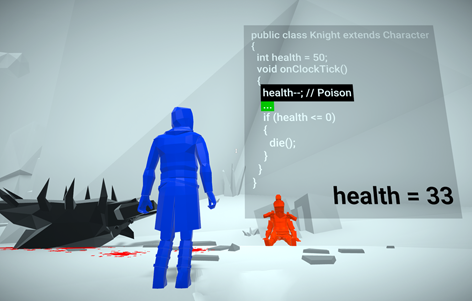}
\end{subfigure}
  \caption{Data type puzzle (L). Curing a wounded knight (R). Placeholders \hlgreen{. . .} indicate where code snippets can be thrown.\protect\footnotemark}
  \label{CodeBreakers}
  \Description{Shows two screenshots from CodeBreakers. First screenshot shows three bugs, each bug associated with a different data type (int, String, boolean) but with a missing value which the player must fill in. Second screenshot shows the code of a knight who is suffering from poison, with a loop which is hurting the knight on every iteration. Player must find the code to stop the poison from killing the knight.}
\end{figure*}
\footnotetext{Note that the avatar model color was changed to gray for this study. See Section \ref{methodssection} for details.\label{footnoteavatarcolorgray}}

\section{Experimental Testbed}

\begin{figure*}
\centering
\begin{subfigure}{.49\textwidth}
  \centering
  \includegraphics[width=1\linewidth]{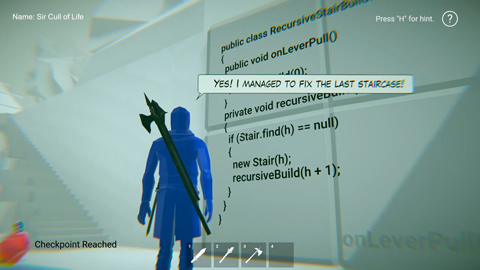}
\end{subfigure}%
\hspace*{\fill}
\begin{subfigure}{.49\textwidth}
  \centering
  \includegraphics[width=1\linewidth]{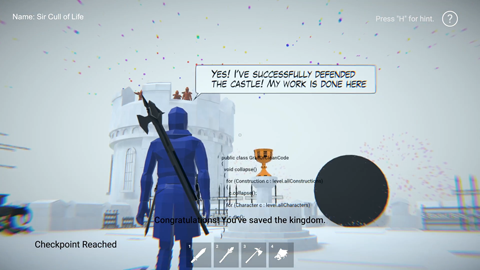}
\end{subfigure}
  \caption{Voice audio occurs in conjunction with speech bubbles that appear on top of the avatar.\footnotesize{\textsuperscript{\ref{footnoteavatarcolorgray}}}}
  \Description{Shows two screenshots from CodeBreakers. Both screenshots have a speech bubble coming from the player and show the situations in which voice audio is played. The first speech bubble says "Yes! I managed the fix the last staircase!" while the second says "Yes! I've successfully defended the castle! My work is done here."}
  \label{CodeBreakersSpeech}
\end{figure*}

Our experimental testbed is CodeBreakers\footnote{Gameplay video: \url{https://youtu.be/x5U-Jd6tKXA}.} \cite{Kao2019}, which was created for conducting avatar-based studies. CodeBreakers is a Java programming game in which players solve increasingly difficult problems by throwing snippets of code. See Figure~\ref{CodeBreakers}. CodeBreakers was iteratively created with feedback from professional game developers, game designers, and Java developers, and it included informal play testing over an eighteen-month span with playtesters. There were 14 total puzzles, spanning 6 levels. CodeBreakers was designed to incorporate best practices on effective learning curves \cite{Linehan2014}. Programming topics include data types, conditionals and control flow, classes and objects, inheritance and interfaces, loops and recursion, and data structures. Each puzzle had up to 3 hints, which are increasingly detailed. Players controlled their character using the keyboard and mouse. CodeBreakers was originally developed for Microsoft Windows and macOS. However, for the purposes of this experiment, CodeBreakers was converted to WebGL and was therefore playable on any PC inside of the browser (e.g., Chrome, Firefox, Safari). See Section \ref{webglconversion} for details. In total, there were 30 possible voice lines that could have been triggered. Other than the first voice line (\textit{What am I doing here? Did my ship crash? How long have I been lying here for? I guess I should get up and look around.}), audio lines typically come before and after each puzzle. For example, prior to puzzle \#7: \textit{The castle is under siege!}. And after completing puzzle \#7: \textit{It worked! I neutralized all of the bugs by using the staff.} These voice lines were accompanied by speech bubbles (see Figure~\ref{CodeBreakersSpeech}).

\section{Methods}\label{methodssection}

For this study, we explicitly aimed to create stereotypically-appearing (and sounding) ``male'' and ``female'' avatars. We created four avatar appearances (two male and two female) and four avatar voices (two male and two female). We made these design decisions with an understanding that a binary view of gender is problematic, but we did so for ecological validity with the majority of existing games. While it would have been possible to create a more inclusive set of gender choices, this might present as a possible confound as such choices are not currently available in most of today's games. Our goal is to develop a baseline understanding of the presence of customization choices that mirror current games. Such baseline understandings can inform future avatar customization research and implementation, in which we hope that more inclusive design choices become the norm. Finally, our rationale for creating two visual choices and two audial choices for each gender was to add a (minimal) degree of visual and audial choice.

\subsection{Model Development}

\begin{figure*}
\centering
\begin{minipage}[t]{.1225\textwidth}
  \centering
  \includegraphics[width=1\linewidth]{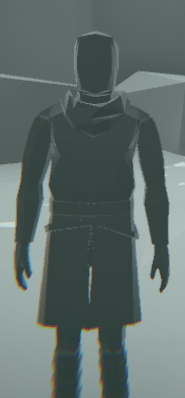}
\end{minipage}%
\begin{minipage}[t]{.1225\textwidth}
  \centering
  \includegraphics[width=1\linewidth]{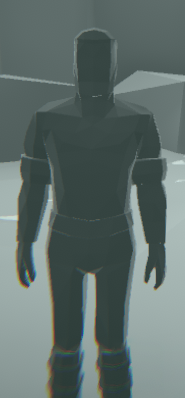}
\end{minipage}%
\begin{minipage}[t]{.1225\textwidth}
  \centering
  \includegraphics[width=1\linewidth]{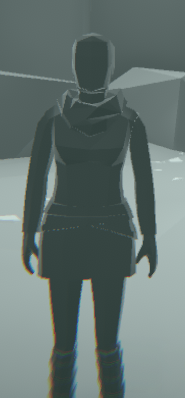}
\end{minipage}%
\begin{minipage}[t]{.1225\textwidth}
  \centering
  \includegraphics[width=1\linewidth]{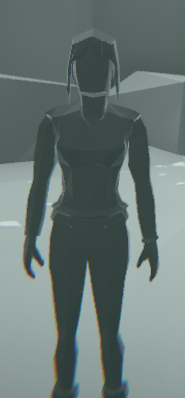}
\end{minipage}%
\hspace*{\fill}
\begin{minipage}[t]{.1225\textwidth}
  \centering
  \includegraphics[width=1\linewidth]{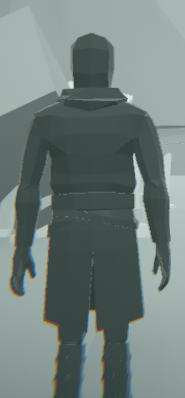}
\end{minipage}%
\begin{minipage}[t]{.1225\textwidth}
  \centering
  \includegraphics[width=1\linewidth]{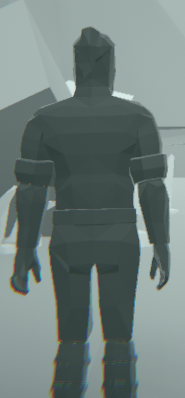}
\end{minipage}%
\begin{minipage}[t]{.1225\textwidth}
  \centering
  \includegraphics[width=1\linewidth]{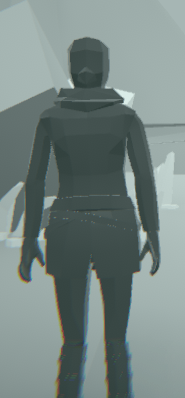}
\end{minipage}%
\begin{minipage}[t]{.1225\textwidth}
  \centering
  \includegraphics[width=1\linewidth]{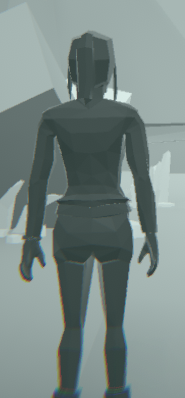}
\end{minipage}
  \caption{Front view (L) and back view (R) of the four models.}
  \label{fig:avatars}
  \Description{Shows the front and back view of the four gray avatars used in the game. All avatars look abstract, without specific facial feature details.}
\end{figure*}

All four models used in this experiment were designed and created from scratch by a professional 3D game artist. The models were purposefully designed to avoid known color effects (e.g., the color red is known to reduce mood, affect, and performance in cognitive-oriented tasks \cite{Gnambs2010,Mehta2008,Meier2015,Hulshof2013,Kuhbandner2013,Kao2016e}). We chose gray because it matched the aesthetic of the game and is not associated with negative physiological effects on cognition and heart rate variability (HF-HRV) \cite{Elliot2011a}. All four models shared the same identical skeleton and joints, and therefore all animations (i.e., idle, walking, picking up code, throwing code, using weapons, falling, dying, stopped in front of a wall, etc.) were identical across the four models. Only visual appearance differed. See Figure~\ref{fig:avatars}.

\subsection{Voice Development}

\subsubsection{Voice Development Goal}\label{voice_dev_goal}

Our goal was to create four avatar voices (two stereotypical male and two stereotypical female). We wanted each voice to be appropriate for the game \textit{and} to be appropriate for either of the two models from the same gender. Additionally, we wanted each male voice to have a ``matching'' female voice as rated on a scale of perceived vocal dimensions---e.g., strong vs. weak, smooth vs. rough, resonant vs. shrill \cite{Gelfer1988}.\footnote{We discuss this scale in more detail in the validation section below.} In other words, we wanted these matched voices to sound as similar as possible. The reason this matching was done was to mitigate confounds from large differences between voices. High variance between voices would add an additional dimension to the manipulation which could influence the study results. Nevertheless, we wanted both male voices to be distinct from one another and both female voices to be distinct from one another. If this were not the case (e.g., both male voices sounded the same), then our manipulation of giving users a \textit{choice} of voice would only be illusory.

\subsubsection{Creating Voices}

We hired two professional voice actors with over ten years of experience in  character voice acting. Both voice actors were screened through their portfolios, which contained samples of their work. Both voice actors provided sample voice clips for CodeBreakers prior to being hired. We decided on hiring two voice actors instead of four because: (1) we could ensure greater overall consistency across voices, helping to bound the variance across voices and (2) both voice actors had demonstrated evidence of being able to perform a multitude of different voices and characters, assurance that each voice actor could produce two unique-sounding voices. Both voice actors self-identified as white and have lived in the U.S. for their entire lives. One voice actor self-identified as male and was 49 years old. The other voice actor self-identified as female and was 38 years old. The two voice actors were instructed to work together to create two ``matching'' voice pairs as described in Section~\ref{voice_dev_goal}. Our goals for the four voices, including the scale of vocal dimensions \cite{Gelfer1988}, were clearly articulated to the voice actors. Additionally, both voice actors familiarized themselves with the game by watching video gameplay of CodeBreakers. Both voice actors were also shown the four models that they were voicing. All voices were recorded in the same professional audio recording studio with both voice actors physically copresent. Identical recording equipment and software was used for recording each voice clip: Sennheiser MK-416 (microphone), Universal Audio Arrow (audio interface), and Ableton Live 10 (digital audio workstation). Completed voice clips were reviewed by the project team, and several iterations were made on the voice clips to ensure that our criteria in Section~\ref{voice_dev_goal} appeared to be satisfied. A total of 120 voice clips (30 per voice) were recorded and finalized. Sample audio clips can be found at \url{https://osf.io/mnpsd/}. M1 is male voice one, M2 is male voice two, F1 is female voice one, and F2 is female voice two.

\subsubsection{Voice Loudness Normalization}

While the same identical recording studio and recording equipment was used for recording each voice, it is possible that relative amplitude (i.e., loudness) could differ between voices, especially between the two different voice actors. To normalize loudness across all voices and voice clips, we adopted the \textit{EBU R 128} (issued by the European Broadcasting Union) standard's recommendation for loudness normalization \cite{ebu2011loudness}. It recommends normalization of audio to -23$\pm$0.5 Loudness Units Full Scale (LUFS), and a max peak of -1 decibel True Peak (dBTP). A professional audio engineer with 15+ years of experience performed this normalization using Nuendo 11 Pro and verified that the loudness normalization recommendation was satisfied.

\subsection{Voice Validation}

\subsubsection{Expert Voice Validation}

\begin{table*}[!htb]\centering\footnotesize
    \begin{tabular}{l@{\hskip12pt}c@{\hskip4pt}c@{\hskip12pt}c@{\hskip4pt}c@{\hskip12pt}c@{\hskip4pt}c@{\hskip12pt}c@{\hskip4pt}c}
        \toprule
        \textsc{Lower Anchor---Upper Anchor} &
        {\textsc{M1}} & {\textsc{(SD)}} &
        {\textsc{M2}} & {\textsc{(SD)}} &
        {\textsc{F1}} & {\textsc{(SD)}} &
        {\textsc{F2}} & {\textsc{(SD)}} \\
        \midrule
        High Pitch---Low Pitch                       & 6.33 & (0.58) & 8.00 & (0.00) & 3.33 & (0.71) & 4.33 & (0.58) \\
        Loud---Soft                                  & 4.67 & (1.53) & 4.67 & (2.31) & 4.67 & (1.41) & 4.00 & (1.73) \\
        Strong---Weak                                & 2.00 & (0.00) & 2.33 & (1.53) & 3.33 & (0.71) & 3.00 & (1.00) \\
        Smooth---Rough                               & 2.33 & (0.58) & 4.00 & (1.73) & 2.00 & (0.00) & 3.67 & (1.15) \\
        Pleasant---Unpleasant                        & 1.67 & (0.58) & 2.33 & (0.58) & 1.67 & (0.71) & 3.00 & (0.00) \\
        Resonant---Shrill                            & 2.67 & (0.58) & 1.67 & (0.58) & 3.67 & (2.83) & 3.33 & (1.15) \\
        Clear---Hoarse                               & 2.33 & (0.58) & 3.67 & (2.89) & 2.33 & (0.71) & 3.67 & (2.89) \\
        Unforced---Strained                          & 3.00 & (1.00) & 4.33 & (2.52) & 3.00 & (0.71) & 3.67 & (1.53) \\
        Soothing---Harsh                             & 3.33 & (0.58) & 2.67 & (0.58) & 2.67 & (0.71) & 3.33 & (1.53) \\
        Melodious---Raspy                            & 3.33 & (0.58) & 4.33 & (2.08) & 2.33 & (0.00) & 4.67 & (0.58) \\
        Breathy Voice---Full Voice                   & 7.00 & (1.73) & 8.33 & (0.58) & 5.00 & (2.83) & 7.00 & (1.00) \\
        Excessively Nasal---Insufficiently Nasal     & 5.00 & (0.00) & 5.00 & (0.00) & 5.00 & (0.00) & 4.00 & (1.00) \\
        Animated---Monotonous                        & 1.67 & (0.58) & 4.67 & (1.53) & 1.67 & (0.00) & 4.00 & (1.73) \\
        Steady---Shaky                               & 2.00 & (0.00) & 2.33 & (0.58) & 2.33 & (0.00) & 2.33 & (0.58) \\
        Young---Old                                  & 4.33 & (0.58) & 5.67 & (0.58) & 3.33 & (0.71) & 4.33 & (1.15) \\
        Slow Rate---Rapid Rate                       & 4.67 & (0.58) & 5.33 & (0.58) & 5.33 & (0.71) & 5.33 & (0.58) \\
        I Like This Voice---I Do Not Like This Voice & 1.67 & (1.15) & 2.00 & (1.00) & 1.67 & (1.41) & 3.33 & (1.53) \\
        \bottomrule
    \end{tabular}

    \vspace{1ex}
    \caption{Mean expert speech pathologist ratings for each voice. All items are rated on a 9-pt Likert scale from 1:Lower Anchor to 9:Upper Anchor.}\label{tab:pathologist}
\end{table*}

To ensure that we had created two distinct matching pairs of voices (similarity \textit{within} each pair but variance \textit{between} them), we hired three expert speech pathologists to evaluate each voice. Each speech pathologist was given instructions to listen to a set of voices then asked to rate each voice on a scale. Each speech pathologist was compensated \$25. Speech pathologists all had at least 10 years of professional speech pathology experience (\textit{M}=20.0, \textit{SD}=8.19), with an average age of \textit{M}=47.67 (\textit{SD}=4.93). Before rating the voices, each speech pathologist was instructed to familiarize themselves with the validated scale on perceptual attributes of voice \cite{Gelfer1988}.\footnote{The scale has been used with speech pathologists revealing modest within-group agreement despite  absence of any training in interpretation of the scale descriptors \cite{Gelfer1988}.} This scale consists of 17 items, and all items are rated on a Likert scale from 1 to 9. Anchor points for each item are listed in Table~\ref{tab:pathologist}. Each speech pathologists was provided the 30 voice clips associated with each voice, and each was asked to listen to the entire set of clips belonging to a single voice before rating that voice. Speech pathologists performed the ratings using their own computers, and they were asked to use the most professional audio equipment available to them to perform the evaluation. Across the three speech pathologists' ratings, we calculated the intraclass correlation to be \textit{ICC}=0.83, 95\% \textit{CI}[0.75, 0.89] (two-way mixed, average measures \cite{Shrout1979}), indicating high agreement. Mean ratings for each voice can be seen in Table~\ref{tab:pathologist}. As a measure of similarity between voices, we then calculated an \textit{absolute mean difference} across the scale between every possible pair of voices. As expected, this difference was lower in the two matched pairs (M1/F1: \textit{M}=0.67; M2/F2: \textit{M}=0.88) when compared to mismatched pairs (M1/F2: \textit{M}=2.33; M2/F1: \textit{M}=1.41) or to same-gender pairs (M1/M2: \textit{M}=1.08; F1/F2: \textit{M}=0.98). Although the same-gender pairs have an absolute mean difference close to the two matched pairs, we attribute some of this due to voice attributes that are oftentimes known to vary naturally between genders (e.g., pitch \cite{Borkowska2011}). Nevertheless, one potential concern arising from these results is that the same-gender voices may not be perceived as distinct from one another. Therefore, we performed an additional crowdsourced validation.

\subsubsection{Crowdsourced Voice Validation}

To ensure that we had created two distinct matching pairs of voices, that all voices would be perceived as as being high quality, that voices would be perceived as the stereotypical intended gender, and that voices across the same gender would be perceived as unique and distinct voices, we ran a crowdsourced validation study. This was to reinforce and extend the prior expert validation. We recruited 91 participants (39\% self-identified as female) on MTurk to rate voices based on sets of audio clips. Each participant was compensated \$1.00 (USD). Participants had a mean age of 40.62 (\textit{SD}=13.82). All participants were from the U.S. After filling out a consent form, each participant was first presented with, randomly, either a stereotypical male or female voice clip of an English word, which they needed to type correctly. This was to ensure that the participant's audio was turned on and working. Each of the following questions was equipped with analytics that tracked the amount of time that each participant spent listening to audio clips. These analytics were used to validate that participants had actually listened to the audio clips before answering the questions. \textasciitilde10\% of participants were removed for not having listened to all audio clips in the study in their entirety.

Participants were then asked to ``\textit{Please listen to \textbf{ALL} of the following audio clips before answering the question below comparing the first (left-side) and second (right-side) voices.}'' And to rate: ``\textit{Besides gender-related voice characteristics, I consider these two voices as similar,}'' on a scale of 1:\textit{Strongly Disagree} to 7:\textit{Strongly Agree}. This question was asked four times comparing the following pairs of voices in a randomized order: M1/F1, M2/F2, M2/F1, and M1/F2. For each comparison, 5 voice clips were selected at random (from the total 30), and those \textit{same} 5 voice clips were shown for both of the two voices being compared (i.e., the same speech dialog).\footnote{Note that randomization is done per participant and per question, so the 5 voice clips selected vary both across questions and across participants.} Results indicated that matched pairs (M1/F1: \textit{M}=5.51, \textit{SD}=1.50; M2/F2: \textit{M}=4.92, \textit{SD}=1.68) were rated to be more highly similar to one another than unmatched pairs (M1/F2: \textit{M}=4.01, \textit{SD}=1.64; M2/F1: \textit{M}=3.13, \textit{SD}=1.71).

Participants were then asked to ``\textit{Please listen to \textbf{ALL} of the following audio clips. All clips belong to one voice. After listening to all of the clips, you will be asked a question regarding the voice.}'' And to rate: ``\textit{Based on the voice you just listened to, please rate the following: The voice is high-quality},'' ``\textit{The speaker sounds (stereotypically) male},'' and ``\textit{The speaker sounds (stereotypically) female}'' on a scale of 1:\textit{Strongly Disagree} to 7:\textit{Strongly Agree}. This question was asked for each of the four voices in randomized order. For each voice, 5 voice clips were selected at random (from the total 30). Results indicated that all voices were perceived to be relatively high quality (M1: \textit{M}=6.02, \textit{SD}=0.80; F1: \textit{M}=6.06, \textit{SD}=0.98;  M2: \textit{M}=5.80, \textit{SD}=1.12; F2: \textit{M}=5.60, \textit{SD}=1.08) and that voices sounded stereotypically male (M1: \textit{M}=6.74, \textit{SD}=0.51; F1: \textit{M}=1.20, \textit{SD}=0.56; M2: \textit{M}=6.85, \textit{SD}=0.39; F2: \textit{M}=1.34, \textit{SD}=0.89) or female (M1: \textit{M}=1.32, \textit{SD}=0.77; F1: \textit{M}=6.79, \textit{SD}=0.44; M2: \textit{M}=1.15, \textit{SD}=0.52; F2: \textit{M}=6.70, \textit{SD}=0.55) as intended.

Participants were then asked to ``\textit{Please listen to \textbf{ALL} of the following audio clips before answering the question below comparing the first (left-side) and second (right-side) voices.}'' And to rate: ``\textit{In comparing the two voices above (left audio clips vs. right audio clips), please rate the following: These two voices are distinct and different from one another,}'' on a scale of 1:\textit{Strongly Disagree} to 7:\textit{Strongly Agree}. This question was asked twice for voices in each gender (M1/M2 and F1/F2) in a random order. For each comparison, 5 voice clips were selected at random. Results indicated that same-gender voice pairs were perceived to be relatively distinct (M1/M2: \textit{M}=5.78, \textit{SD}=1.07; F1/F2: \textit{M}=5.73, \textit{SD}=1.30). Participants then entered demographic information.

\subsection{Model and Voice Integration}

\subsubsection{WebGL Conversion and Technical Testing}\label{webglconversion}

\begin{figure*}
\centering
\begin{subfigure}{.49\textwidth}
  \centering
  \includegraphics[width=\linewidth]{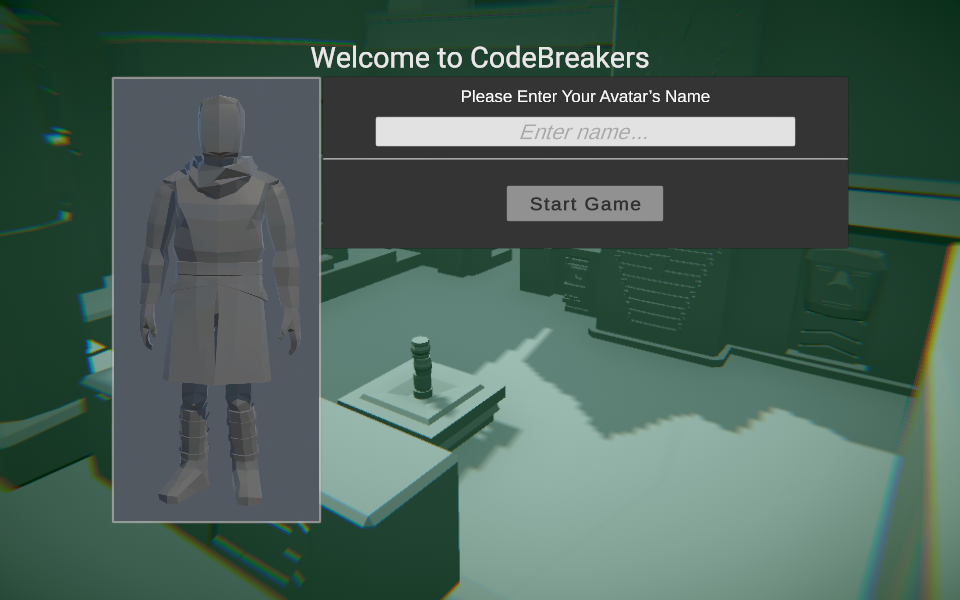}
  \caption{\footnotesize\textit{Choice-None}: Participant is randomly assigned both model and voice.}
\end{subfigure}
\hspace*{\fill}
\begin{subfigure}{.49\textwidth}
  \centering
  \includegraphics[width=\linewidth]{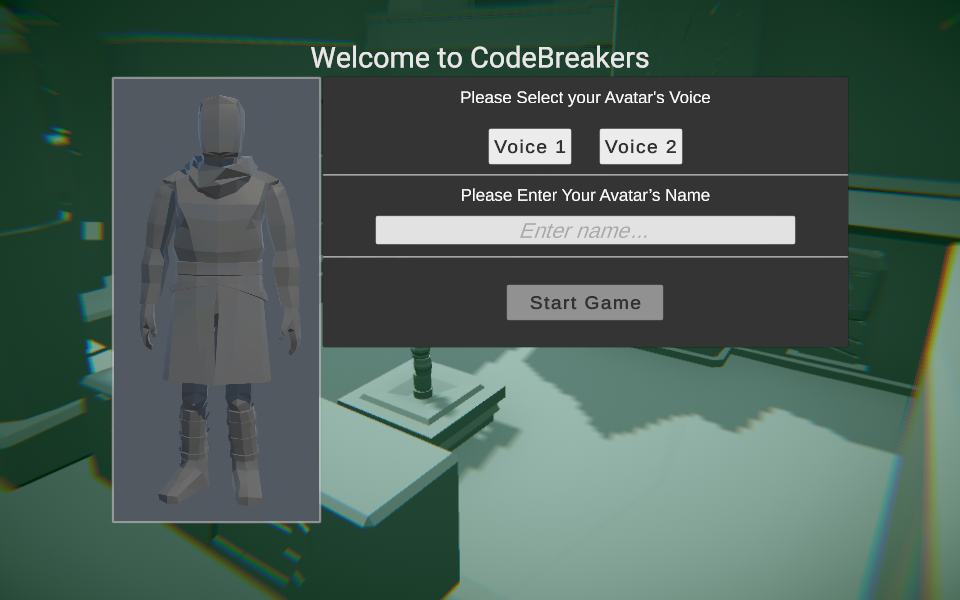}
  \caption{\footnotesize\textit{Choice-Audio}: Participant is randomly assigned model and chooses voice.}
\end{subfigure}
\\ [1.6ex]
\begin{subfigure}{.49\textwidth}
  \centering
  \includegraphics[width=\linewidth]{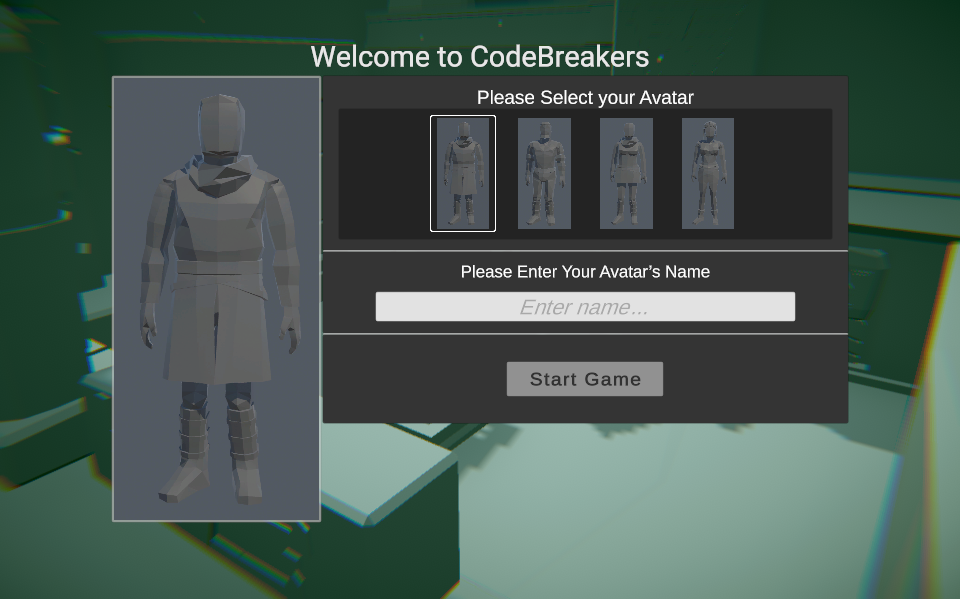}
  \caption{\footnotesize\textit{Choice-Visual}: Participant chooses model and is randomly assigned voice.}
\end{subfigure}
\hspace*{\fill}
\begin{subfigure}{.49\textwidth}
  \centering
  \includegraphics[width=\linewidth]{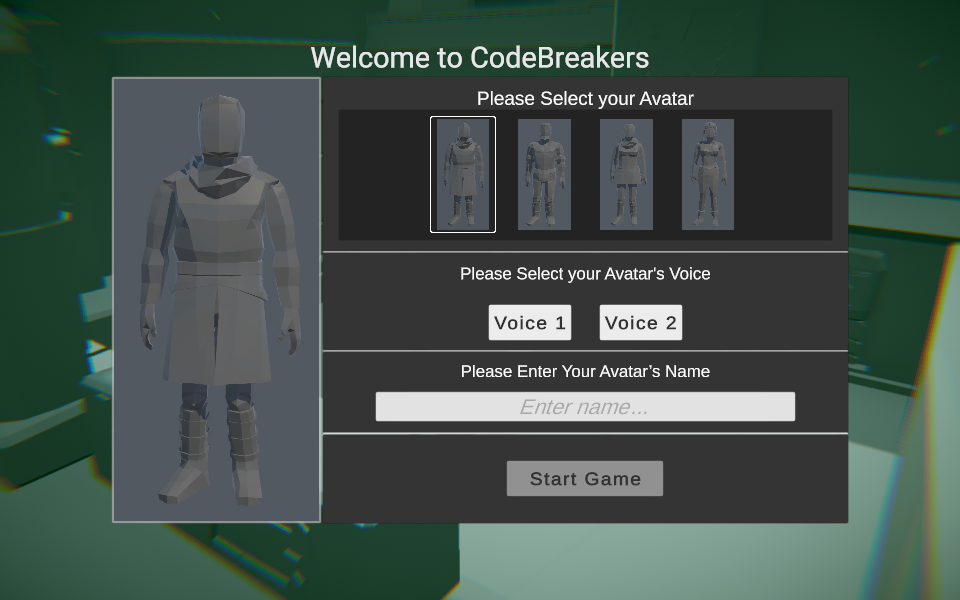}
  \caption{\footnotesize\textit{Choice-All}: Participant chooses both model and voice.}
\end{subfigure}
  \caption{Avatar customization screens.}
  \label{fig:conditions}
  \Description{Shows four screenshots, one of each condition. (a) Choice-None: Participant is randomly assigned both model and voice. (b) Choice-Audio: Participant is randomly assigned model and chooses voice. (c) Choice-Visual: Participant chooses a model and is randomly assigned voice. (d) Choice-All: Participant chooses both model and voice.}
\end{figure*}

Over 4 months, the original CodeBreakers game, which is playable on machines running either Microsoft Windows or macOS \cite{kao2021chiplayavatarvoice}, was converted to WebGL to allow for a more convenient play experience. The WebGL version is playable on any PC inside of the browser (e.g., Chrome, Firefox, Safari). This conversion was performed by a professional game development team with expertise in game optimization. During the conversion process, we iterated on the game internally every few days and externally every few weeks. Our main goal during these iterations was to ensure that performance (e.g., frames per second) was adequate and that there were no technical issues (e.g., crashing). Internal iterations were performed by the development and research team where feedback was fed into the next iteration. Performance profiling tools were used extensively to diagnose areas of the game (e.g., code loops, rendering of certain geometry) responsible for increased CPU and memory usage. External iterations were performed when we wanted the game to be tested more widely. We performed iterations with batches of 10-20 participants at a time on MTurk. Participants were asked to play the \textit{entire} game and were provided a walkthrough video in case they were unable to progress. This ensured that each participant would cover the breadth of the entire game. Data, including gameplay metrics, performance, crash logs, and PC details, was automatically logged on the server for further analysis. Participants could report any issues, problems, or concerns they experienced during playtesting. A total of 121 participants, all from the U.S., took part in external playtesting. Each participant was compensated \$10 (USD). Our testing ended when no new technical issues arose in the most recent internal and external iterations, all known technical issues were fixed, and the game performed adequately (e.g., frames per second, load times) under a wide variety of PCs. Additionally, the development and research team agreed that, for all intents and purposes, the WebGL game played and felt identical to the original.
\begin{figure*}
  \centering
\includegraphics{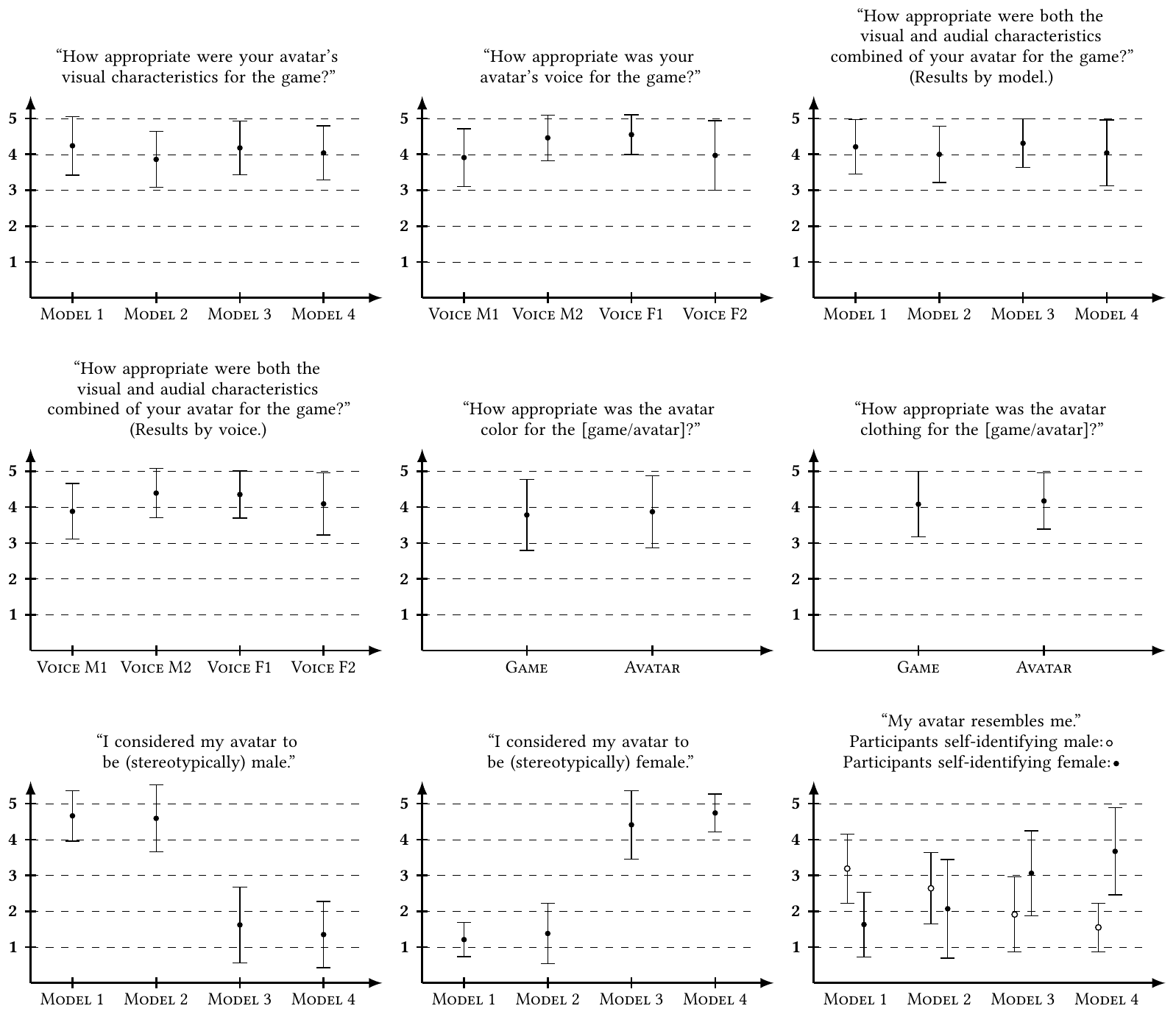}
  \caption{Model and voice validation summary graphs. Error bars show $\pm$SD.}\label{fig:manipcheckgraphs}
  \Description{Model and voice validation summary graphs. Error bars show ±SD.}
\end{figure*}

\subsubsection{Character Customization UI}\label{sec:charactercustomizationui}

A professional game UI designer created four different character customization screens that we requested. These also correspond to our experimental conditions. (See Figure~\ref{fig:conditions}.) We made the explicit design decision never to allow mismatched model--voice gender pairings (i.e., male model and female voice or vice versa), since this may be unnatural for players, lacks general ecological validity with existing games, and may be an experimental confound (e.g., in conditions where one or both features are assigned at random). Therefore, avatar customization is, in all cases, a two-step process that involves first choosing or being assigned a model (one of four), then choosing or being assigned a voice (one of two since the model has already been selected, and there are only two voices corresponding to the designed stereotypical gender).

In \textit{Choice-None}, the player does not have any choice over the model or voice. Both model and voice are randomly assigned. In \textit{Choice-Audio}, a model is randomly preselected, and a player is able to choose the voice. In \textit{Choice-Visual}, the player chooses a model, after which the voice is randomly assigned. In \textit{Choice-All}, the player chooses both model and voice. Note that the two voices corresponding to ``Voice 1'' and ``Voice 2'' will differ depending on the model selected. In \textit{Choice-All}, both voice options are grayed out and unavailable until a model has been selected. If a different model is selected after a voice has been selected, the voice is automatically deselected. In all conditions, players must enter a name for their character. For conditions that allow for a model choice (\textit{Choice-Visual} and \textit{Choice-All}), the UI initially shows an empty box where the selected model would normally appear (i.e., no model is selected by default). For conditions that allow for a voice choice (\textit{Choice-Audio} and \textit{Choice-All}), no voice is selected by default (i.e., one of the two voices must be selected manually by the player). When a voice is selected, a single audio clip is played from that voice so that players can compare voices. In all conditions, players must complete all customization options available (e.g., name, model, voice) before the ``Start Game'' button becomes available. Character customization conditions were designed in this manner to minimize differences between conditions, while still varying the manipulations (visual choice and audial choice).

\subsubsection{Expert UI Validation}

To assess the appropriateness of our character customization UIs, we performed a validation study with three professional game UI designers. Game UI designers were recruited from the online freelancing platform Upwork, and were each paid \$20 (USD). The job posting was \textit{Assess Character Customization Interface in Educational Game}, and the job description stated that we were looking for expert game UI designers to evaluate a set of character customization interfaces in an educational game. The three UI designers had an average of 9.00 (\textit{SD}=4.36) years of UI design experience and an average of 7.67 (\textit{SD}=5.69) years of game development experience. UI designers all had work experience and portfolios that reflected recent UI design and game development experience (all within one year). UI designers were instructed to give their honest opinions and were told their responses would be anonymous, and proceeded to our survey. Each UI designer was first asked to watch 30 minutes of gameplay footage from CodeBreakers to familiarize themselves with the game. Afterwards, each designer loaded CodeBreakers WebGL on their own machine and interacted with every version of the UI in a randomized order. After interacting with a specific version of the UI, the UI designer was asked to rate ``\textit{The character customization interface is appropriate for the game},'' on a scale of 1:\textit{Strongly Disagree} to 7:\textit{Strongly Agree}. UI designers were asked to rate each interface individually, \textit{not} in comparison to the other interfaces they had already seen. UI designers were also able to report open-ended feedback. The survey took approximately 1.5 hours to complete. Responses showed that UI designers generally agreed that the character customization interface was appropriate (\textit{Choice-None}: \textit{M}=6.67, \textit{SD}=0.58; \textit{Choice-Audial}: \textit{M}=6.33, \textit{SD}=1.16; \textit{Choice-Visual}: \textit{M}=6.33, \textit{SD}=1.16; \textit{Choice-All}: \textit{M}=7.00, \textit{SD}=0.00). One UI designer did note as open-ended feedback that they had not expected to be able to choose a voice for their character since this is not a commonly available feature in games, but it was stated that this did not play a role in the designer's ratings.

\subsubsection{Model and Voice Integration Validation}

To assess whether the models and voices that we had developed would be perceived as appropriate for the game, we recruited 120 participants (43\% female) on MTurk. All 120 participants played CodeBreakers using the \textit{Choice-None} condition (i.e., randomly assigned model and voice). Participants played the game for a minimum of 5 minutes, but they were allowed to play as long as they liked beyond the 5-minute mark. Random assignments were roughly even across models (24.2\%/24.2\%/32.5\%/19.2\%) and voices (24.2\%/27.5\%/26.7\%/21.7\%). For the remainder of this section, ratings described for models follow the left-to-right order of models shown in Figure~\ref{fig:avatars}. See Figure~\ref{fig:manipcheckgraphs} for graphs summarizing the validation results.\footnote{All validation questions are found in the graphs except for ``\textit{How appropriate was the avatar design overall?}'' for which summary statistics are provided in the text.}

To assess whether models overall visually fit the game, we asked, ``\textit{How appropriate were your avatar's visual characteristics for the game?}'' on a scale from 1:\textit{Inappropriate} to 5:\textit{Appropriate}. Scores tended between neutral and appropriate for each model (\textit{M}=4.24, \textit{SD}=0.83; \textit{M}=3.86, \textit{SD}=0.79; \textit{M}=4.18, \textit{SD}=0.76; \textit{M}=4.04, \textit{SD}=0.77). To assess whether voices overall audially fit the game, we asked ``\textit{How appropriate was your avatar's voice for the game?}'' on a scale from 1:\textit{Inappropriate} to 5:\textit{Appropriate}. Scores again tended between neutral and appropriate for each voice (M1: \textit{M}=3.91, \textit{SD}=0.82; M2: \textit{M}=4.46, \textit{SD}=0.65; F1: \textit{M}=4.55, \textit{SD}=0.57; F2: \textit{M}=3.97, \textit{SD}=0.98). To assess whether models \textit{and} voices in combination fit the game, we asked, ``\textit{How appropriate were both the visual and audial characteristics combined of your avatar for the game?}'' on a scale from 1:\textit{Inappropriate} to 5:\textit{Appropriate}. Scores again tended between neutral and appropriate for each model (\textit{M}=4.21, \textit{SD}=0.77; \textit{M}=4.00, \textit{SD}=0.80; \textit{M}=4.31, \textit{SD}=0.69; \textit{M}=4.04, \textit{SD}=0.93) and for each voice (M1: \textit{M}=3.88, \textit{SD}=0.79; M2: \textit{M}=4.39, \textit{SD}=0.70; F1: \textit{M}=4.35, \textit{SD}=0.67; F2: \textit{M}=4.09, \textit{SD}=0.88). To assess whether models' individual visual features (color and clothing) were appropriate for the avatar, and for the game, we asked, ``\textit{How appropriate was the avatar color for the game?}'', ``\textit{How appropriate was the avatar color for the avatar?}'', ``\textit{How appropriate was the avatar clothing for the game?}'', ``\textit{How appropriate was the avatar clothing for the avatar?}'', and ``\textit{How appropriate was the avatar design overall?}'', on a scale from 1:\textit{Inappropriate} to 5:\textit{Appropriate}. Overall scores were between neutral and appropriate for appropriateness of avatar color (Game: \textit{M}=3.78, \textit{SD}=1.00; Avatar: \textit{M}=3.87, \textit{SD}=1.02), avatar clothing (Game: \textit{M}=4.08, \textit{SD}=0.93; Avatar: \textit{M}=4.17, \textit{SD}=0.80), and avatar design overall (\textit{M}=4.06, \textit{SD}=0.87). To assess whether models were perceived as the stereotypical gender we had designed them to be, we asked, ``\textit{I considered my avatar to be (stereotypically) male,}'' and ``\textit{I considered my avatar to be (stereotypically) female,}'' on a scale from 1:\textit{Strongly Disagree} to 5:\textit{Strongly Agree}. Participants rated the models designed to be stereotypically male as male (\textit{M}=4.66, \textit{SD}=0.72; \textit{M}=4.59, \textit{SD}=0.95; \textit{M}=1.62, \textit{SD}=1.07; \textit{M}=1.35, \textit{SD}=0.94) and models designed to be stereotypically female as female (\textit{M}=1.21, \textit{SD}=0.49; \textit{M}=1.38, \textit{SD}=0.86; \textit{M}=4.41, \textit{SD}=0.97; \textit{M}=4.74, \textit{SD}=0.54). To assess whether avatars bore a visual similarity with players, we asked participants to rate ``\textit{My avatar resembles me,}'' on a scale from 1:\textit{Strongly Disagree} to 5:\textit{Strongly Agree}. Participants who self-identified as male had scores tending towards neutral for male models (\textit{M}=3.19, \textit{SD}=0.98; \textit{M}=2.64, \textit{SD}=1.01; \textit{M}=1.91, \textit{SD}=1.06; \textit{M}=1.55, \textit{SD}=0.69) while participants who self-identified as female had scores tending towards neutral for female models (\textit{M}=1.63, \textit{SD}=0.92; \textit{M}=2.07, \textit{SD}=1.39; \textit{M}=3.06, \textit{SD}=1.20; \textit{M}=3.67, \textit{SD}=1.23). As expected, participants in general did not find close visual similarity with their avatars (likely in part due to their abstract design), with some natural variation across avatars and gender.

\subsection{Study Preregistration}

Our study was preregistered on the Open Science Framework (OSF). Hypotheses, exploratory analyses, experiment design, data collection, sample size, and measures are contained in our preregistration.\footnote{Preregistration: \url{https://osf.io/dbvkp/}.\\ Raw Data: \url{https://osf.io/mnpsd/}.}

\subsection{Conditions}

The study uses a 2 x 2 factorial design. We manipulate visual choice (choice vs. assignment) and audial choice (choice vs. assignment). The manipulations are as follows:
\begin{itemize}[noitemsep,topsep=3pt]
\item \textbf{Choice-None}: Participant is randomly assigned both model and voice.
\item \textbf{Choice-Audio}: Participant is randomly assigned model and chooses voice.
\item \textbf{Choice-Visual}: Participant chooses a model and is randomly assigned voice.
\item \textbf{Choice-All}: Participant chooses both model and voice.
\end{itemize}

The only difference between each of these conditions is the character customization interface that appeared at the beginning of the game, which manipulated choice vs. assignment for model and voice. See Figure~\ref{fig:conditions} and Section~\ref{sec:charactercustomizationui} for details on how the character customization interface was implemented in these different conditions. All other aspects of the experiment were identical across conditions.

\subsection{Measures}\label{sec:measures}

In line with best practices on measurement reporting, we report \textit{what} we are measuring, \textit{how} we are measuring, and \textit{why} are we measuring in this way \cite{aeschbach2021systematic}.

\subsubsection{Avatar Identification (Player Identification Scale)}

Avatar identification is a ``temporary alteration of media users' self-concept through adoption of perceived characteristics of a media person'' \cite{christophVideo2009}. For measuring avatar identification, we use the player identification scale (PIS) \cite{vanlooyPlayer2012}. The PIS measures three dimensions of avatar identification on a 5-pt Likert scale (1:\textit{Strongly Disagree} to 5:\textit{Strongly Agree}): similarity identification (e.g., ``My character is similar to me''), embodied identification (e.g., ``In the game, it is as if I become one with my character''), and wishful identification (e.g., ``I would like to be more like my character''). We use the PIS as it has been validated \cite{vanlooyPlayer2012} and is used extensively in the HCI literature on avatars---e.g., \cite{birkFostering2016}.

\subsubsection{Autonomy (Player Experience of Need Satisfaction)}

Autonomy is the sense that one has volition and is doing activities for interest and personal value \cite{Ryan2006}. We use the PENS scale \cite{Ryan2006} to measure autonomy on a 7-pt Likert scale (1:\textit{Do Not Agree} to 7:\textit{Strongly Agree})---e.g., ``The game provides me with interesting options and choices.'' We use the PENS autonomy subscale as it has been empirically validated on multiple occasions---e.g., \cite{Johnson2018}.

\subsubsection{Intrinsic Motivation (Intrinsic Motivation Inventory)}

Intrinsic motivation is one's willingness to engage in an activity because the activity is satisfying in and of itself \cite{ryan2000intrinsic}. The subscale of the IMI, interest/enjoyment, is the primary measure of intrinsic motivation used in the research literature \cite{ryan2000intrinsic}. This is due to both interest and enjoyment being strong contributors to intrinsic motivation \cite{reeve1989interest}. Items are rated on a 7-pt Likert scale (1:\textit{Not At All True} to 7:\textit{Very True})---e.g., ``I enjoyed doing this activity very much.'' We chose to use the IMI to measure intrinsic motivation since it is well validated \cite{mcauley1989psychometric}.

\enlargethispage{-20pt}

\subsubsection{Immersion (Player Experience Inventory)}

Immersion is \textit{a sense of immersion and cognitive absorption, experienced by the player} \cite{Abeele2020}. We use the Player Experience Inventory (PXI) to measure immersion, which uses three items to measure immersion on a 7-pt Likert scale, from -3:\textit{Strongly Disagree} to +3:\textit{Strongly Agree}---e.g., ``I was no longer aware of my surroundings while I was playing.'' We use the PXI immersion subscale, since it has been extensively validated and was designed specifically for games user research \cite{Abeele2020}.

\subsubsection{Motivated Behavior (Time Played)}

We operationalize motivated behavior as the time spent playing the game. Time on task is a behavioral measure that has been linked to motivation \cite{Sansone1992,ryan1991ego} and is an objective measure of motivation in this study. Note that in the current study, participants are \textit{required} to play at least 10 minutes, after which playing longer is optional.

\subsubsection{Motivation For Future Play and Likelihood of Game Recommendation}

Both motivation for future play and likelihood of game recommendation are measured using questions identical to a previous study \cite{Peng2012}. Specifically, motivation for future play was measured using three items based on Ryan, Rigby, and Przybylski \cite{Ryan2006}: ``Given the chance I would play this game in my free time,'' ``I would like to spend more time playing this game,'' and ``I would like to continue playing this game'' \cite{Peng2012}. Participants rated the three items on a 7-pt Likert scale from 1:\textit{Strongly Disagree} to 7:\textit{Strongly Agree}. Likelihood of game recommendation was assessed using the question ``How likely would you be to recommend this game to others?'' on a 7-pt Likert scale from 1:\textit{Extremely Unlikely} to 7:\textit{Extremely Likely} \cite{Peng2012}. These measures allow us to understand how willing a player is to come back to a game, and how willing a player is to recommend the game to others. Both of these measures have been frequently used in the literature---e.g., PENS autonomy has been shown to positively predict both motivation for future play and likelihood of game recommendation in prior studies \cite{Peng2012,Ryan2006,przybylski2009motivating}. Motivation for future play showed good reliability, $\alpha$=0.98.

\subsection{Sample Size Determination}

To calculate a priori sample size, we perform two separate sample size determination calculations (both of these are specified in our preregistration at \url{https://osf.io/dbvkp/}). The first calculation is based on a 2 x 2 ANOVA for testing H1 and H2. G*Power 3.1 was used to perform this calculation using an effect size of small (0.1), $\alpha$=0.05, and 95\% power. G*Power 3.1 found that a sample size of \textit{N}=1302 would be required \cite{lakens2013calculating}.


For H3, H4, H5, H6, and H7, our sample size calculation is based on moderated mediation analyses. We performed Monte Carlo simulations in R. We first specify the complete model (i.e., containing X, M1, M2, M3, M4, W, and Y with the appropriate relationships) using the \textit{lavaan} package, with parameter estimates of 0.1 (e.g., correlations between variables). We then use the \textit{simsem} package to create Monte Carlo simulations using 1000 bootstraps.\footnote{This is considered well above the number of iterations needed: \url{https://kb.palisade.com/index.php?pg=kb.page&id=125}.} These simulations provided estimations of statistical power for each path, from which we use the lowest power value from all paths as the cutoff. We modified sample size iteratively ($\pm$10) until the necessary power was reached. We performed 10 simulations to confirm that a specified sample size would reach the desired minimum power. The random number generator's seed was re-randomized for every simulation. These Monte Carlo simulations determined that, for a power of 95\% and a confidence level of 95\%, a sample size of 1500 would be necessary.\footnote{The sample size calculation takes a similar approach to Schoemann, Boulton, and Short \cite{Schoemann2017}.} Therefore, to ensure the necessary power across both sample size determinations (\textit{N}=1302 and \textit{N}=1500), we use \textit{N}=1500.

\begin{figure*}
\centering
\begin{subfigure}{.49\textwidth}
 \centering
 \includegraphics[width=1\linewidth]{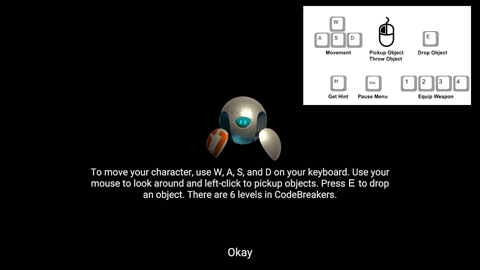}
 \caption{The robotic agent introduces the game.}
\end{subfigure}
\hspace*{\fill}
\begin{subfigure}{.49\textwidth}
 \centering
 \includegraphics[width=1\linewidth]{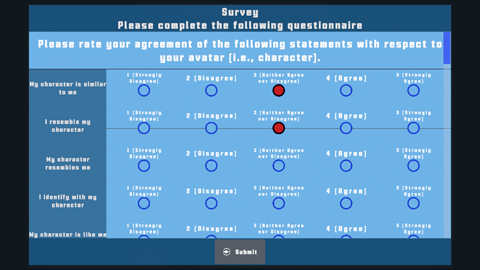}
 \caption{Survey that players complete after 10 minutes of gameplay.}
 \label{fig:survey}
\end{subfigure}
  \caption{Screenshots from the experiment.}
  \label{fig:harley}
  \Description{Two screenshots from the experiment. One image shows a robotic agent introducing the game to the player, along with controls for playing the game. The other image shows the in-game survey that players complete after 10 minutes of gameplay.}
\end{figure*}

\subsection{Participants}

We recruited 1527 (47.6\% female, 1.2\% gender variant, 0.4\% transgender) participants with an average age of \textit{M}=37.26 (\textit{SD}=11.14) from MTurk.\footnote{Note that we explicitly recruited a slightly larger number than we had calculated (\textit{N}=1500) in case of loss of data during data screening.} Workers on MTurk complete Human Intelligence Tasks (HITs), including research experiments. Studies show that MTurk provides data of similar quality \cite{buhrmester2011amazon}, diversity \cite{chandler2016conducting,horton2011online,berinsky2012evaluating}, and reliability \cite{Mason2012,buhrmester2011amazon} as typical samples (e.g., college students). Participants were each paid \$5.00 (USD). The HIT was available to workers in the U.S. over the age of 18 who had a computer with working audio. For quality control, workers were required to have a HIT approval rate
>95\%. The Purdue University Institutional \nobreak Review Board (IRB) approved the study. All participants were asked to provide informed consent.

\subsubsection{Data Screening}
We screened all participants' responses. Specifically, we carefully screened participants' who had at least three survey measures with zero variance (excluding likelihood of game recommendation, since this was only a single question) or with $\pm$3SD. A fairly large number of respondents met the criteria of at least three survey measures with zero variance (\textasciitilde40\%),\footnote{This large number is not unexpected given that most survey measures are measuring a single concept---e.g., immersion, autonomy, and low variance is expected within these individual survey measures. Nevertheless, it is important to manually inspect these responses for data quality---e.g., ``straight-liners'' that always pick the same answer option \cite{bruhlmann2018surveys}.} and these responses were scrutinized further (e.g., reverse-coded items and open-ended questions). All responses were deemed legitimate, except for one respondent who responded to all questions (including reverse-coded items) with the same answer. This respondent was removed from further analysis (\textit{N}=1526 remaining participants).

\subsubsection{Experience With Video Games and Programming}
Participants reported playing an average of \textit{M}=8.5 (\textit{SD}=10.5) hours of video games per week, approximately matching the global average of \textit{M}=8.45 \cite{stateofonlinegaming2021}. On a scale from 1:\textit{Minimal} to 7:\textit{Extensive}, participants rated their prior experience playing video games (``How would you rate your prior experience playing video games?'') as \textit{M}=4.72 (\textit{SD}=1.81) and their prior programming experience (``How would you rate your prior programming experience?'') as \textit{M}=2.34 (\textit{SD}=1.64). Next, we adapted several questions on programming experience from \cite{Siegmund2014}. On a scale from 1:\textit{Very Inexperienced} to 5:\textit{Very Experienced}, participants rated their programming experience compared to experts (``How do you estimate your programming experience compared to experts with 20 years of practical experience?'') as \textit{M}=1.34 (\textit{SD}=0.85), their programming experience compared to beginners (``How do you estimate your programming experience compared to beginner programmers?'') as \textit{M}=2.05 (\textit{SD}=1.18), their programming experience in Java specifically (``How experienced are you with the Java programming language?'') as \textit{M}=1.56 (\textit{SD}=0.93), and their experience with an object-oriented paradigm (``How experienced are you with the object-oriented programming paradigm?'') as \textit{M}=1.72 (\textit{SD}=1.13). Therefore, our sample contains participants who are regularly exposed to video games and have low prior programming experience. ANOVAs found that there were no significant differences between conditions on prior gaming experience ($F[3, 1522]$=0.422, $p$=0.737, $\eta_{p}^{2}$=0.001), programming experience ($F[3, 1522]$=0.264, $p$=0.851, $\eta_{p}^{2}$=0.001), and Java programming experience ($F[3, 1522]$=0.263, $p$=0.852, $\eta_{p}^{2}$=0.001).

\subsection{Design}

A between-subjects factorial design was used. Each participant was randomly assigned to one of four possible conditions. Participant counts in each condition were approximately equal (\textit{M}=381.5, \textit{SD}=5.8).

\subsection{Procedure}

Participants first filled out an IRB-approved consent form. Participants were informed that they could exit the game at any time after playing 10 minutes. Participants then began playing CodeBreakers. At the beginning of the game, participants underwent an audio check during which they were required to type a spoken English word. Participants then used the avatar customization interface corresponding to their condition. A robotic agent then engaged in a short conversation with the player. The robot was animated with audio dialogue generated through an automatic voice generator \cite{LingoJam}. After a brief introduction, the participant was provided instructions on how to play the game. See Figure~\ref{fig:harley}a. Participants were told they could exit the game at any time after playing 10 minutes by pressing ESC on their keyboard, then clicking quit game. The participant then began playing the game. During gameplay, the text ``\textit{Time Remaining for Survey}'' appeared at the top of the screen, with a countdown timer starting from 10 minutes. Once the 10 minutes had elapsed, participants were automatically presented an in-game survey which contained the PIS, PENS autonomy, IMI interest/enjoyment, PXI immersion, motivation for future play, and likelihood of game recommendation questions. See Figure~\ref{fig:harley}b. All participant game data was automatically logged (e.g., time played, avatar customization choices). After the survey was completed, a message box appeared, reminding participants that they could now quit at any time, and that they could continue playing for as long as they liked. The message at the top of the game screen which had shown the time remaining was replaced by the message ``\textit{You may play for as long as you like and quit at any time by pressing ESC and clicking Quit Game}.'' Once participants quit the game (or completed all 6 levels), participants were then asked to describe in their own words any problems encountered. Participants then filled out a set of questions about prior video game experience, programming experience, and demographics.

\subsection{Analysis}

Data was analyzed using SPSS 23 and the PROCESS macro for SPSS \cite{hayes2017introduction}. Factorial 2 x 2 ANOVAs were used to study the effects of visual choice and audial choice on the PIS (H1), and PENS autonomy (H2).\footnote{ANOVAs are considered robust to non-normality, especially at larger sample sizes \cite{Blanca2017}.} We then performed a parallel mediation analysis with visual choice (\textit{X}), similarity identification (\textit{M1}), embodied identification (\textit{M2}), wishful identification (\textit{M3}), autonomy (\textit{M4}), and IMI interest/enjoyment (\textit{Y}) (H3). We used PROCESS model 4 \cite{hayes2017introduction}. The parallel mediation was repeated using the different outcomes of interest (\textit{Y}): PXI immersion (H4); time spent playing (H5); motivation for future play (H6); and likelihood of game recommendation (H7). In order to perform exploratory analyses on whether audial choice (\textit{W}) moderates paths (direct and indirect) between \textit{X} and \textit{Y}, we used PROCESS model 59. We used an $\alpha$ of 0.05. These analyses were all preregistered at \url{https://osf.io/dbvkp/}. 

\section{Results}

\subsection{Checking For Model- and Voice-Specific Effects}

To ensure that there were no effects of a specific model, or a specific voice, on collected measures (see Section~\ref{sec:measures} for our measures), we used one-way MANOVA. First, we grouped all participants who were assigned a model \textit{randomly}---i.e., participants in the \textit{Choice-None} and \textit{Choice-Audio} conditions. Second, we created another group of participants who were assigned a voice randomly---i.e., participants in the \textit{Choice-None} and \textit{Choice-Visual} conditions. We only chose participants who were assigned an avatar or voice randomly (and not through choice), since this gives the best approximation of how an avatar or voice may influence a player while avoiding the confound of a self-selection effect. Using the two groups, we then ran two MANOVAs with the IVs of either avatar (group 1) or voice (group 2) and the DVs of our collected measures. Prior to running our MANOVAs, we checked both assumption of homogeneity of variance and homogeneity of covariance by the test of Levene's Test of Equality of Error Variances and Box's Test of Equality of Covariance Matrices; and both assumptions were met by the data ($p$>0.05 for Levene's, and $p$>0.001 for Box's). However, Levene's test was violated for the measure of \textit{time played} in the MANOVA for group 2 ($p$<0.05). To deal with this violation, we used the more conservative Pillai's Trace \cite{tabachnick2007using}. We also set the more conservative significance criterion of $p$<0.01 (two-tailed) for univariate testing as suggested in the literature \cite{tabachnick2007using}. There was no statistically significant difference in our measures based on model, $F[27,2194]$=0.89, $p$=0.630, Wilk's $\Lambda$=0.969, $\eta_{p}^{2}$=0.011. There was no statistically significant difference in our measures based on voice, $F[27, 2229]$=1.241, $p$=0.183, Pillai's Trace=0.044, $\eta_{p}^{2}$=0.015. Therefore, when assigned randomly, neither a specific model nor a specific voice had a significant effect on our measures.

\subsection{H1: Effect of Manipulation on Avatar Identification}

\begin{table*}[!htb]\centering\footnotesize\setlength{\tabcolsep}{3pt}
    \newcommand{\ttvbf}[1]{\multicolumn{2}{c}{\bfseries#1}}
    \newcommand{\ttvrm}[1]{\multicolumn{2}{c}{#1}}
    \newcommand{\phs}{\phantom{<}}
    \newcommand{\myhead}[1]{\multicolumn{2}{>{\centering\arraybackslash}p{0.12\linewidth-2\tabcolsep}}{#1}}
    \newcommand{\mysubh}{
        \multicolumn{1}{>{\centering\arraybackslash}p{0.06\linewidth-2\tabcolsep}}{$M$}&
        \multicolumn{1}{>{\centering\arraybackslash}p{0.06\linewidth-2\tabcolsep}}{$SD$}
        }
    \begin{tabular}{lcccccccc}
    \toprule
    & \myhead{PIS Similarity} & \myhead{PIS Embodied} & \myhead{PIS Wishful} & \myhead{PENS Autonomy} \\
    \cmidrule(l{2pt}r{2pt}){2-3} \cmidrule(l{2pt}r{2pt}){4-5} \cmidrule(l{2pt}r{2pt}){6-7} \cmidrule(l{2pt}r{2pt}){8-9}
    &
    \mysubh & \mysubh & \mysubh & \mysubh \\
    \midrule
    \multicolumn{9}{l}{\textsc{Visual No Choice}}\\
    \textsc{Audial No Choice} & 2.59 & 1.08 & 2.89 & 1.15 & 2.50 & 1.04 & 4.04 & 1.70 \\
    \textsc{Audial Choice}    & 2.46 & 0.98 & 2.75 & 1.14 & 2.34 & 1.00 & 3.80 & 1.75 \\
     & \\ [-5pt]
    \hdashline
     & \\ [-4pt]
    \multicolumn{9}{l}{\textsc{Visual Choice}}\\
    \textsc{Audial No Choice} & 2.91 & 1.08 & 3.04 & 1.20 & 2.67 & 1.10 & 4.19 & 1.73 \\
    \textsc{Audial Choice}    & 3.21 & 1.04 & 3.35 & 1.10 & 2.96 & 1.07 & 4.48 & 1.60 \\
    \midrule
    \multicolumn{9}{l}{\textsc{Main Effect Choice Visual}}\\
    $F$            & \ttvbf{98.719}    & \ttvbf{40.104}    & \ttvbf{52.538}    & \ttvbf{23.017}    \\
    $p$            & \ttvbf{<0.001}    & \ttvbf{<0.001}    & \ttvbf{<0.001}    & \ttvbf{<0.001}    \\
    $\eta_{p}^{2}$ & \ttvbf{\phs0.061} & \ttvbf{\phs0.026} & \ttvbf{\phs0.033} & \ttvbf{\phs0.015} \\
    \midrule
    \multicolumn{9}{l}{\textsc{Main Effect Choice Audial}}\\
    $F$            & \ttvrm{2.449}  & \ttvrm{2.041}  & \ttvrm{1.354}          & \ttvrm{0.089}  \\
    $p$            & \ttvrm{0.118}  & \ttvrm{0.153}  & \ttvrm{0.245}          & \ttvrm{0.765}  \\
    $\eta_{p}^{2}$ & \ttvrm{0.002}  & \ttvrm{0.001}  & \ttvrm{0.001}          & \ttvrm{0.000}  \\
    \midrule
    \multicolumn{9}{l}{\textsc{Interaction Effect}}\\
    $F$            & \ttvbf{15.561}    & \ttvbf{14.564}    & \ttvbf{16.998}    & \ttvbf{9.356} \\
    $p$            & \ttvbf{<0.001}    & \ttvbf{<0.001}    & \ttvbf{<0.001}    & \ttvbf{0.002} \\
    $\eta_{p}^{2}$ & \ttvbf{\phs0.010} & \ttvbf{\phs0.009} & \ttvbf{\phs0.011} & \ttvbf{0.006} \\
    \midrule
    \multicolumn{9}{l}{\textsc{Simple Effect Choice Audial (Visual No Choice)}}\\
    $F$            & \ttvrm{2.832} & \ttvrm{2.850} & \ttvrm{4.378}                             & \ttvrm{3.808} \\
    $p$            & \ttvrm{0.093} & \ttvrm{0.092} & \ttvrm{\phantom{$\dagger$}0.037$\dagger$} & \ttvrm{0.051} \\
    $\eta_{p}^{2}$ & \ttvrm{0.002} & \ttvrm{0.002} & \ttvrm{0.003}                             & \ttvrm{0.002} \\
    \midrule
    \multicolumn{9}{l}{\textsc{Simple Effect Choice Audial (Visual Choice)}}\\
    $F$            & \ttvbf{15.178}    & \ttvbf{13.756}    & \ttvbf{13.974}    & \ttvbf{5.638} \\
    $p$            & \ttvbf{<0.001}    & \ttvbf{<0.001}    & \ttvbf{<0.001}    & \ttvbf{0.018} \\
    $\eta_{p}^{2}$ & \ttvbf{\phs0.010} & \ttvbf{\phs0.009} & \ttvbf{\phs0.009} & \ttvbf{0.004} \\
    \bottomrule
    \end{tabular}\\

    \vspace{1ex}
    {\footnotesize Visual Choice \textit{df}=1, Audial Choice \textit{df}=1, Interaction \textit{df}=1, Error \textit{df}=1522}\\
    {\footnotesize $\dagger$Not significant due to Bonferroni-adjusted $\alpha$=0.025 for simple effect}\\

    \vspace{0ex}
    \caption{Results for effects of visual choice and audial choice on PIS (H1) and PENS autonomy (H2). Significant results are bold.}\label{tab:h1h2}
\end{table*}

From Table~\ref{tab:h1h2}, factorial 2 x 2 ANOVAs (choice visual x choice audial) found main effects of choice visual on similarity identification, embodied identification, and wishful identification (H1.1 supported). In contrast, there were no main effects of choice audial (H1.2 not supported). However, a significant interaction effect was found between choice visual and choice audial on similarity identification, embodied identification, and wishful identification (H1.3 not supported). Significant interaction effects were further probed through a simple effects analysis. As this involved two additional tests, the significance threshold was Bonferroni-adjusted to $p$=0.025. Simple effects analysis found that in all cases, the effect of choice audial when there was \textit{no} visual choice was not significant. However, in all cases, the effect of choice audial when there \textit{was} visual choice was significant and positive. Therefore, in the absence of a visual avatar choice, choice of avatar voice has no effect, but in the presence of a visual avatar choice, choice of avatar voice has a significantly positive effect on similarity identification, embodied identification, and wishful identification. Effect sizes ($\eta_{p}^{2}$) are in the small-to-medium (0.01 to 0.09) range for main effects of choice visual, and small (0.01) for interaction effects.\footnote{Small effect sizes are not uncommon in games user research due to the complexity of player-game interactions \cite{birkFostering2016,Steinemann2015,Birk2015,Zendle2015}.}

\subsection{H2: Effect of Manipulation on Autonomy}

From Table~\ref{tab:h1h2}, a factorial 2 x 2 ANOVA (choice visual x choice audial) found a main effect of choice visual on autonomy (H2.1 supported). In contrast, there was no main effect of choice audial (H2.2 not supported). However, a significant interaction effect was found between choice visual and choice audial on autonomy (H1.3 not supported). The significant interaction effect was further probed through a simple effect analysis. As this involved two additional tests, the significance threshold was Bonferroni-adjusted to $p$=0.025. The simple effect analysis found that the effect of choice audial when there was \textit{no} visual choice was not significant. However, the effect of choice audial when there \textit{was} visual choice was significant and positive. Therefore, in the absence of a visual avatar choice, choice of avatar voice has no effect, but in the presence of a visual avatar choice, choice of avatar voice has a significantly positive effect on autonomy. The effect size ($\eta_{p}^{2}$) is considered small.

\begin{figure*}[!htb]\centering
    \newlength{\myboxo}\setlength{\myboxo}{33pt}
    \begin{tikzpicture}[x=130pt,y=40pt,box/.style={rectangle,draw,text width=55pt,minimum height=\myboxo,align=center},every node/.style={font=\footnotesize},>=latex]
        \node[box] (L) at (-1,0) {Visual Avatar\\ Choice\\ (X)};
        \coordinate (L1) at ([yshift=0.5\myboxo]L.east);
        \coordinate (L2) at ([yshift=0.25\myboxo]L.east);
        \coordinate (L3) at ([yshift=-0.25\myboxo]L.east);
        \coordinate (L4) at ([yshift=-0.5\myboxo]L.east);
        \node[box] (M1) at (0,1.7) {Similarity\\ Identification\\ (M$_1$)};
        \node[box] (M2) at (0,0.7) {Embodied\\ Identification\\ (M$_2$)};
        \node[box] (M3) at (0,-0.7) {Wishful\\ Identification\\ (M$_3$)};
        \node[box] (M4) at (0,-1.7) {Autonomy\\ (M$_4$)};
        \node[box] (R) at (1,0) {Outcome\\ (Y)};
        \coordinate (R1) at ([yshift=0.5\myboxo]R.west);
        \coordinate (R2) at ([yshift=0.25\myboxo]R.west);
        \coordinate (R3) at ([yshift=-0.25\myboxo]R.west);
        \coordinate (R4) at ([yshift=-0.5\myboxo]R.west);
        \foreach \n in {1,2,3,4} {
            \draw[->] (L\n.east)--(M\n.west) node[midway,above,inner sep=2pt,circle] {$a_\n$};
            \draw[->] (M\n.east)--(R\n.west) node[midway,above,inner sep=2pt,circle] {$b_\n$};
        }
        \draw[->] (L.east)--(R.west) node[midway,above] {$c'$};
    \end{tikzpicture}
    \caption{Mediation model being tested for H3 through H7.}
    \label{fig:process4}
\end{figure*}
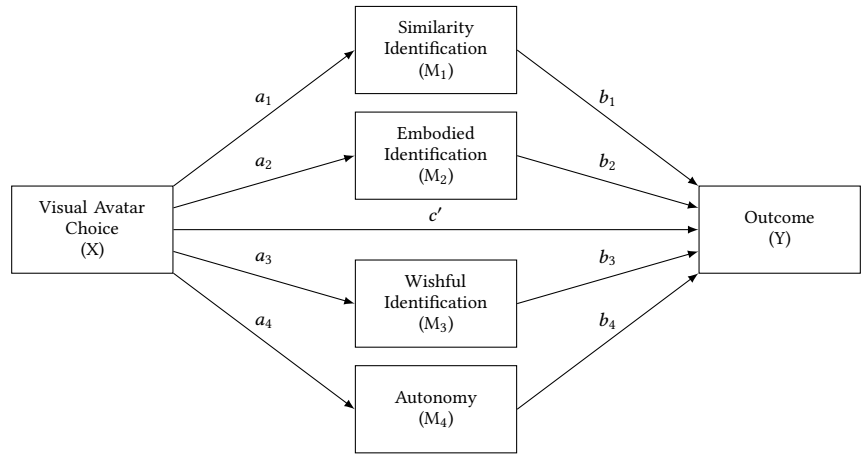
\subsection{H3--H7: Mediation and Moderation Analyses}

\begin{table*}[!htb]\centering\footnotesize
    \newlength{\squished}
    \setlength{\squished}{1pt}
\tabcolsep2pt
\begin{tabular}{ccc} 
    \begin{tabular}{c@{\hspace{0.4cm}}rr@{}l}
        \multicolumn{4}{c}{\textsc{Similarity Identification}} \\
        \toprule
        $a_1$ & \multicolumn{1}{c}{$b_1$} & \multicolumn{2}{c}{$a_1b_1$} \\
        \midrule
        \multicolumn{4}{l}{\textsc{Intrinsic Motivation}}              \\[\squished]
        {\textbf{0.535}\text{***}} & {0.033}                  & {0.018;\ }         & \textit{CI}[$-0.027, 0.065$] \\
        \midrule
        \multicolumn{4}{l}{\textsc{Immersion}}                         \\[\squished]
        {\textbf{0.535}\text{***}} & {$-0.034$}                 & {$-0.018$;\ }        & \textit{CI}[$-0.067, 0.030$] \\
        \midrule
        \multicolumn{4}{l}{\textsc{Time Spent Playing}}                \\[\squished]
        {\textbf{0.535}\text{***}} & {\textbf{31.65}\text{*}} & {\textbf{16.94;}}\ \ & \textbf{\textit{CI}[2.091, 33.02]} \\
        \midrule
        \multicolumn{4}{l}{\textsc{Motivation for Future Play}}        \\[\squished]
        {\textbf{0.535}\text{***}} & {0.025}                  & {0.013;\ }         & \textit{CI}[$-0.050, 0.075$] \\
        \midrule
        \multicolumn{4}{l}{\textsc{Likelihood of Game Recommendation}} \\[\squished]
        {\textbf{0.535}\text{***}} & {0.084}                  & {0.045;\ }         & \textit{CI}[$-0.013, 0.105$] \\
                \bottomrule
    \end{tabular}
    \hspace{2ex}%
    \begin{tabular}{c@{\hspace{0.4cm}}rr@{}l}
        \multicolumn{4}{c}{\textsc{Embodied Identification}} \\
        \toprule
        $a_2$                        & \multicolumn{1}{c}{$b_2$}                        & \multicolumn{2}{c}{$a_2b_2$}                                      \\
        \midrule
\\[\squished]
        {\textbf{0.375}\text{***}} & {\textbf{0.264}\text{***}} & {\textbf{0.099;}}\ \ & \textbf{\textit{CI}[0.057, 0.148]} \\
        \midrule
\\[\squished]
        {\textbf{0.375}\text{***}} & {\textbf{0.573}\text{***}} & {\textbf{0.215;}}\ \ & \textbf{\textit{CI}[0.144, 0.293]} \\
        \midrule
\\[\squished]
        {\textbf{0.375}\text{***}} & {$-1.693$}                   & {$-0.635$;\ } & {\textit{CI}[$-10.24, 9.037]$}        \\
        \midrule
\\[\squished]
        {\textbf{0.375}\text{***}} & {\textbf{0.296}\text{***}} & {\textbf{0.111;}}\ \ & \textbf{\textit{CI}[0.062, 0.169]} \\
        \midrule
\\[\squished]
        {\textbf{0.375}\text{***}} & {\textbf{0.187}\text{***}} & {\textbf{0.070;}}\ \ & \textbf{\textit{CI}[0.028, 0.120]} \\
        \bottomrule
    \end{tabular}
    \hspace{2ex}%
    \begin{tabular}{c@{\hspace{0.4cm}}rr@{}l}
        \multicolumn{4}{c}{\textsc{Wishful Identification}} \\
        \toprule
        $a_3$                        & \multicolumn{1}{c}{$b_3$}                        & \multicolumn{2}{c}{$a_3b_3$}                                         \\
        \midrule
\\[\squished]
        {\textbf{0.393}\text{***}} & {\textbf{$-$0.085}\text{*}}  & {\textbf{$-$0.033;}}\ \ & \textbf{\textit{CI}[$-$0.068, $-$0.001]} \\
        \midrule
\\[\squished]
        {\textbf{0.393}\text{***}} & {$-$0.059}                   & {$-0.023;$}\ \ & {\textit{CI}[$-0.061, 0.010$]}           \\
        \midrule
\\[\squished]
        {\textbf{0.393}\text{***}} & {\textbf{33.10}\text{*}}   & {\textbf{13.02;}}\ \ & \textbf{\textit{CI}[2.376, 25.48]}    \\
        \midrule
\\[\squished]
        {\textbf{0.393}\text{***}} & {\textbf{0.158}\text{**}}  & {\textbf{0.062;}}\ \ & \textbf{\textit{CI}[0.016, 0.117]}    \\
        \midrule
\\[\squished]
        {\textbf{0.393}\text{***}} & {\textbf{0.174}\text{***}} & {\textbf{0.069;}}\ \ & \textbf{\textit{CI}[0.023, 0.122]}    \\
        \bottomrule
    \end{tabular}
\\ \\
\vspace{2ex}
    \begin{tabular}{c@{\hspace{0.4cm}}rr@{}l}
        \multicolumn{4}{c}{\textsc{Autonomy}} \\
        \toprule
        $a_4$                        & \multicolumn{1}{c}{$b_4$}                        & \multicolumn{2}{c}{$a_4b_4$} \\
        \midrule
        \multicolumn{4}{l}{\textsc{Intrinsic Motivation}}              \\[\squished]
        {\textbf{0.420}\text{***}} & {\textbf{0.686}\text{***}} & \hspace{2em}{\textbf{0.288;}}\ \ &  \textbf{\textit{CI}[0.171, 0.409]} \\
        \midrule
        \multicolumn{4}{l}{\textsc{Immersion}}                         \\[\squished]
        {\textbf{0.420}\text{***}} & {\textbf{0.298}\text{***}} & {\textbf{0.125;}}\ \ &  \textbf{\textit{CI}[0.073, 0.182]} \\
        \midrule
        \multicolumn{4}{l}{\textsc{Time Spent Playing}}                \\[\squished]
        {\textbf{0.420}\text{***}} & {$-0.913$}                   & {$-0.384;$}\ \         &  {\textit{CI}[$-5.683, 4.548$]}       \\
        \midrule
        \multicolumn{4}{l}{\textsc{Motivation for Future Play}}        \\[\squished]
        {\textbf{0.420}\text{***}} & {\textbf{0.670}\text{***}} & {\textbf{0.281;}}\ \ &  \textbf{\textit{CI}[0.165, 0.400]} \\
        \midrule
        \multicolumn{4}{l}{\textsc{Likelihood of Game Recommendation}} \\[\squished]
        {\textbf{0.420}\text{***}} & {\textbf{0.706}\text{***}} & {\textbf{0.297;}}\ \ &  \textbf{\textit{CI}[0.176, 0.422]} \\
        \bottomrule
    \end{tabular}
    \hspace{2ex}%
    \begin{tabular}{c}
            \multicolumn{1}{c}{\textsc{Direct Effect}}\\
        \toprule
            $c'$ \\
       \midrule
        %
\\[\squished]
        {0.015} \\
        \midrule
\\[\squished]
        {0.048} \\
        \midrule
\\[\squished]
        {\textbf{41.66}\text{*}} \\
        \midrule
\\[\squished]
        {0.053} \\
        \midrule
\\[\squished]
        {0.030} \\
        \bottomrule
    \end{tabular}
    \hspace{2ex}%
    \begin{tabular}{c}
        \multicolumn{1}{c}{\textsc{Total Effect}}\\
        \toprule
        $c$ \\
        \midrule
        %
\\[\squished]
        \textbf{0.387}\text{***} \\
        \midrule
\\[\squished]
        \textbf{0.347}\text{***} \\
        \midrule
\\[\squished]
        \textbf{70.60}\text{***} \\
        \midrule
\\[\squished]
        \textbf{0.521}\text{***} \\
        \midrule
\\[\squished]
        \textbf{0.510}\text{***} \\
        \bottomrule
    \end{tabular}\\
\multicolumn{3}{c}{\footnotesize * significant at $p < 0.05$; ** significant at $p < 0.01$; *** significant at $p < 0.005$; significant $a_xb_x$ based on 95\% \textit{CI}.}
\end{tabular}
    \vspace{1ex}
    \caption{Mediation results with visual avatar choice ($X$), similarity identification ($M_1$), embodied identification ($M_2$), wishful identification ($M_3$), autonomy ($M_4$), and each outcome variable ($Y$). Regression coefficients $a_x$~($X$$\rightarrow$$M_x$), $b_x$~($M_x$$\rightarrow$$Y$), $c'$~(direct $X$$\rightarrow$$Y$), $c$~(total $X$$\rightarrow$$Y$), and $a_xb_x$. All presented effects are unstandardized. Significant results are bold.}
    \label{tab:mediation}
    \Description{Mediation model being tested for H3 through H7. Shows a parallel mediation model with Visual Avatar Choice (X), Similarity Identification (M1), Embodied Identification (M2), Wishful Identification (M3), Autonomy (M4), and Outcome (Y).}
\end{table*}

\subsubsection{Assumption Checks}

Mediation analyses require several important assumptions to be met \cite{berry1993understanding}: (1) linearity, (2) normality, (3) homoscedasticity, (4) absence of strong multicollinearity, and (5) absence of extreme outliers. (1) To ensure linearity, we plotted scatterplots between each predictor variable and dependent variable. All such permutations were plotted and manually checked to ensure the linearity assumption was satisfied; bivariate correlations were also tested \cite{berry1993understanding}. Linearity was found to be satisfied in all cases. (2) We used PROCESS' bootstrapping option, which makes no assumptions about the distribution of the underlying data \cite{hayes2017introduction}. Therefore, the normality assumption is automatically satisfied. (3) We used robust standard errors (HC4 \cite{Cribari-Neto2004}) in all of our analyses, automatically satisfying the assumption of homoscedasticity \cite{hayes2017introduction}. (4) To ensure absence of strong multicollinearity, we verified the VIF (Variance Inflation Factor) between all predictor variables and dependent variables. A VIF > 5 is generally a cause for concern, while a VIF > 10 indicates a serious collinearity problem \cite{menard2002applied}. All VIF scores were below 5, satisfying the assumption of absence of strong multicollinearity. (5) To ensure absence of extreme outliers, we performed outlier testing. The only variable that is at risk of outliers is \textit{time played} (our independent variables are binary and cannot contain outliers by design; similarly, Likert-scale data do not contain outliers). However, outlier testing requires a normal distribution.\footnote{At a more theoretical level, this is because a distribution must be assumed in order to be able to classify a data point as lying outside the expected range.} A Kolmogorov-Smirnov test ($p$<0.05), a Shapiro-Wilk test ($p$<0.05), and a Q-Q plot all indicated that the variable \textit{time played} does not meet the assumption of normality.\footnote{This was an expected result by design. Because our experimental design was to have a minimum playtime of 10 minutes, we expected a right-skewed distribution with a peak at the 10 minute mark (and no participants below 10 minutes), making the data non-normal.} Therefore, we first perform the data transformation described by Templeton \cite{templeton2011two}. This is a two-step process: (i) transformation into a percentile rank; and (ii) an inverse-normal transformation. After this process, a Kolmogorov-Smirnov test ($p$=0.200), a Shapiro-Wilk test ($p$=0.997), and a Q-Q plot all indicated that the transformed variable was normally distributed. We then used an Interquartile Range (IQR) multiplier of 2.2 for outlier detection \cite{Hoaglin1987}, and we found no outliers. Therefore, our mediation analysis assumptions are met.

\subsubsection{Hypothesis Tests}\label{sec:hypothesistests}

\begin{figure*}[!htb]\centering
    \newlength{\myboxa}\setlength{\myboxa}{32pt}
    \begin{tikzpicture}[x=130pt,y=40pt,box/.style={rectangle,draw,text width=55pt,minimum height=\myboxa,align=center},every node/.style={font=\footnotesize},>=latex]
        \node[box] (L) at (-1,0) {Visual Avatar\\ Choice\\ (X)};
        \coordinate (L1) at ([yshift=0.5\myboxa]L.east);
        \coordinate (L2) at ([yshift=0.25\myboxa]L.east);
        \coordinate (L3) at ([yshift=-0.25\myboxa]L.east);
        \coordinate (L4) at ([yshift=-0.5\myboxa]L.east);
        \node[box] (M1) at (0,1.5) {Similarity\\ Identification\\ (M$_1$)};
        \node[box] (M2) at (0,0.5) {Embodied\\ Identification\\ (M$_2$)};
        \node[box] (M3) at (0,-0.5) {Wishful\\ Identification\\ (M$_3$)};
        \node[box] (M4) at (0,-1.5) {Autonomy\\ (M$_4$)};
        \node[box] (M5) at (0,-2.5) {Audial Avatar\\ Choice\\ (W)};
        \node[box] (R) at (1,0) {Outcome\\ (Y)};
        \coordinate (R1) at ([yshift=0.5\myboxa]R.west);
        \coordinate (R2) at ([yshift=0.25\myboxa]R.west);
        \coordinate (R3) at ([yshift=-0.25\myboxa]R.west);
        \coordinate (R4) at ([yshift=-0.5\myboxa]R.west);
        \foreach \n in {1,2,3,4} {
            \draw[->] (L\n.east)--(M\n.west) coordinate[pos=0.5](Pl1\n) coordinate[pos=0.6](Pl2\n) coordinate[pos=0.7](Pl3\n) coordinate[pos=0.8](Pl4\n);
            \draw[->] (M\n.east)--(R\n.west) coordinate[pos=0.5](Pr1\n) coordinate[pos=0.4](Pr2\n) coordinate[pos=0.3](Pr3\n) coordinate[pos=0.2](Pr4\n) coordinate[pos=0.1](Pr5\n);
        }
        \draw[->] (L.east)--(R.west);

        \draw[->] ([yshift=-0.25\myboxa]M5.west) -| (Pl14);
        \draw[->] ([yshift=-0.25\myboxa]M5.west) -| (Pl23);
        \draw[->] (M5.west) -| (Pl32);
        \draw[->] (M5.west) -| (Pl41);

        \draw[->] ([yshift=-0.25\myboxa]M5.east) -| (Pr14);
        \draw[->] ([yshift=-0.25\myboxa]M5.east) -| (Pr23);
        \draw[->] (M5.east) -| (Pr32);
        \draw[->] (M5.east) -| (Pr41);
        \draw[->] ([yshift=0.25\myboxa]M5.east) -| (Pr51|-L.east);

    \end{tikzpicture}
  \caption{Moderated mediation model being tested in our exploratory analyses.}
  \label{fig:process59}
  \Description{Moderated mediation model being tested in our exploratory analyses. Audial Avatar Choice (W) as potentially moderating the paths in Figure 7.}
\end{figure*}
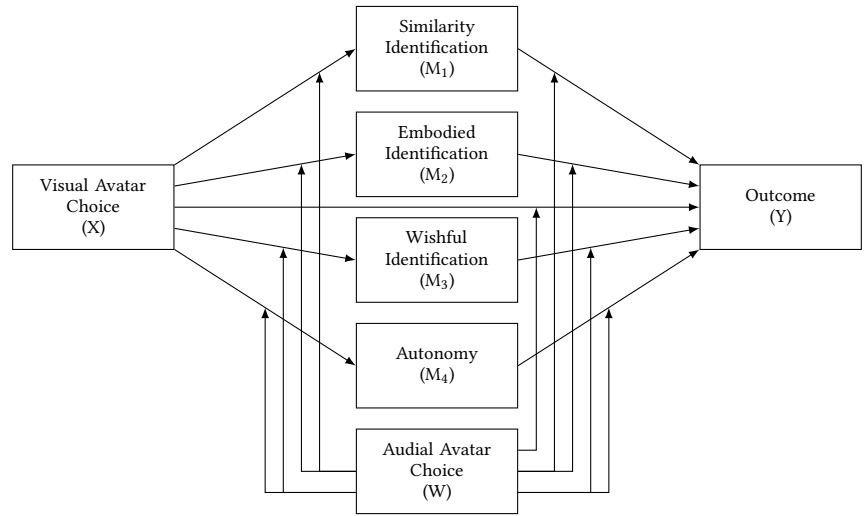

\begin{table*}[!htb]\centering\footnotesize
    \newcommand{\myheadt}[1]{\multicolumn{2}{>{\centering\arraybackslash}p{0.17\linewidth-2\tabcolsep}}{\textsc{#1}}}
    \newcommand{\mysubht}{
        \multicolumn{1}{>{\centering\arraybackslash}p{0.085\linewidth-2\tabcolsep}}{$M$}&
        \multicolumn{1}{>{\centering\arraybackslash}p{0.085\linewidth-2\tabcolsep}}{$SD$}
        }
\begin{tabular}{lrrrrrrrr}
    \toprule
    &
    \multicolumn{4}{c}{\textsc{Visual No Choice}} &
    \multicolumn{4}{c}{\textsc{Visual Choice}} \\
    \cmidrule(l{2pt}r{2pt}){2-5} \cmidrule(l{2pt}r{2pt}){6-9}
    &
    \myheadt{Audial No Choice} &
    \myheadt{Audial Choice}    &
    \myheadt{Audial No Choice} &
    \myheadt{Audial Choice}    \\ \cmidrule(l{2pt}r{2pt}){2-3} \cmidrule(l{2pt}r{2pt}){4-5} \cmidrule(l{2pt}r{2pt}){6-7} \cmidrule(l{2pt}r{2pt}){8-9}
    \textsc{Variable} & \mysubht & \mysubht & \mysubht & \mysubht \\
    \midrule
    Intrinsic Motivation              & 4.57\hspace{1em}   & 1.70\hspace{1em}   & 4.35\hspace{1em}   & 1.76\hspace{1em}   & 4.72\hspace{1em}   & 1.70\hspace{1em}   & 4.97\hspace{1em}   & 1.47\hspace{1em}   \\
    Immersion                         & 0.77\hspace{1em}   & 1.44\hspace{1em}   & 0.58\hspace{1em}   & 1.59\hspace{1em}   & 0.89\hspace{1em}   & 1.44\hspace{1em}   & 1.15\hspace{1em}   & 1.39\hspace{1em}   \\
    Time Spent Playing                & 864.57\hspace{1em} & 319.14\hspace{1em} & 844.52\hspace{1em} & 292.78\hspace{1em} & 905.35\hspace{1em} & 370.15\hspace{1em} & 944.18\hspace{1em} & 412.18\hspace{1em} \\
    Motivation for Future Play        & 3.87\hspace{1em}   & 1.95\hspace{1em}   & 3.65\hspace{1em}   & 2.01\hspace{1em}   & 4.10\hspace{1em}   & 2.02\hspace{1em}   & 4.46\hspace{1em}   & 1.96\hspace{1em}   \\
    Likelihood of Game Recommendation & 3.94\hspace{1em}   & 1.94\hspace{1em}   & 3.67\hspace{1em}   & 2.00\hspace{1em}   & 4.12\hspace{1em}   & 1.97\hspace{1em}   & 4.50\hspace{1em}   & 1.96\hspace{1em}   \\
    \bottomrule
    \end{tabular}

    \vspace{1ex}
    \caption{Descriptives for outcomes in H3 through H7. Immersion was on a Likert scale from  -3 to +3. Intrinsic Motivation, Motivation for Future Play, and Likelihood of Game Recommendation were on Likert scales from 1 to 7.}\label{tab:descriptivesh3toh7}
\end{table*}

\looseness-1The mediation model being tested can be seen in Figure~\ref{fig:process4}. From Table~\ref{tab:mediation}, we can see that visual choice has a direct effect ($c'$) on time spent playing only (H5.1 supported; H3.1, H4.1, H6.1, and H7.1 not supported). A 95\% bias-corrected confidence interval based on 10,000 bootstrap samples indicates several significant indirect effects on intrinsic motivation ($a_2b_2$ and $a_3b_3$\footnote{Note the negative coefficient, meaning that higher wishful identification is related to \textit{lower} intrinsic motivation, which was unexpected. \textit{All} other significant effects in our model were positive coefficients.} supporting H3.2, $a_4b_4$ supporting H3.3), immersion ($a_2b_2$ supporting H4.2, $a_4b_4$ supporting H4.3), time spent playing ($a_1b_1$ and $a_3b_3$ supporting H5.2, H5.3 not supported), motivation for future play ($a_2b_2$ and $a_3b_3$ supporting H6.2, $a_4b_4$ supporting H6.3), and likelihood of game recommendation ($a_2b_2$ and $a_3b_3$ supporting H7.2, $a_4b_4$ supporting H7.3). Therefore, we conclude that visual choice \textit{directly} affects time spent playing, and \textit{indirectly} affects intrinsic motivation (via embodied identification, wishful identification, and autonomy), immersion (via embodied identification and autonomy), time spent playing (via similarity identification and wishful identification), motivation for future play (via embodied identification, wishful identification, and autonomy), and likelihood of game recommendation (via embodied identification, wishful identification, and autonomy). Descriptives for each variable can be seen in Table~\ref{tab:descriptivesh3toh7}.

\subsubsection{Exploratory Analyses}

\begin{table*}[!htb]\footnotesize\setlength{\tabcolsep}{4pt}
    \resizebox{\textwidth}{!}{%
    \begin{tabular}{llr@{}lr@{}lr@{}lr@{}lr@{}l}
    \toprule
    \textsc{Variable} & &
    \multicolumn{2}{c}{\textsc{$X \rightarrow M_1 \rightarrow Y$}} &
    \multicolumn{2}{c}{\textsc{$X \rightarrow M_2 \rightarrow Y$}} &
    \multicolumn{2}{c}{\textsc{$X \rightarrow M_3 \rightarrow Y$}} &
    \multicolumn{2}{c}{\textsc{$X \rightarrow M_4 \rightarrow Y$}} &
    \multicolumn{2}{c}{\textsc{$X \rightarrow Y$}} \\
    \midrule
    Intrinsic Mot. &  Index of MM    & $-0.080$;\ \ & CI[$-0.179, 0.018$]                  & \textbf{0.132};\ \ & \textbf{\textit{CI}[0.038, 0.233]} & 0.003;\ \           & CI[$-0.071, 0.077$]                  & \textbf{0.364};\ \ & \textbf{\textit{CI}[0.128, 0.600]} & \multicolumn{2}{c}{---} \\
                   &  Effect at AC$=0$ & 0.036;\ \          & CI[$-0.003, 0.084$]                  & 0.036;\ \          & CI[$-0.006, 0.089$]                  & $-0.022$;\ \         & CI[$-0.054, 0.001$]                  & 0.104;\ \          & CI[$-0.063, 0.272$]                  & $-0.006$;\ \                  &  CI[$-0.153, 0.142$]                  \\
                   &  Effect at AC$=$1 & $-0.044$;\ \         & CI[$-0.132, 0.047$]                  & \textbf{0.168};\ \ & \textbf{\textit{CI}[0.092, 0.259]} & $-0.019$;\ \        & CI[$-0.087, 0.049$]                  & \textbf{0.468};\ \ & \textbf{\textit{CI}[0.302, 0.637]} & 0.045;\ \                  &  CI[$-0.099, 0.190$]                  \\
                   \\ [-1ex]
    Immersion      &  Index of MM    & 0.012;\ \          & CI[$-0.098, 0.130$]                  & \textbf{0.255};\ \ & \textbf{\textit{CI}[0.104, 0.418]} & $-0.013$;\ \         & CI[$-0.098, 0.069$]                  & \textbf{0.172};\ \ & \textbf{\textit{CI}[0.062, 0.290]} & \multicolumn{2}{c}{---} \\
                   &  Effect at AC$=0$ & $-0.019$;\ \        & CI[$-0.063, 0.018$]                  & 0.086;\ \         & CI[$-0.012, 0.188$]                  & $-0.013$;\ \        & CI[$-0.044, 0.006$]                  & 0.042;\ \         & CI[$-0.026, 0.113$]                  & 0.026;\ \                   &  CI[$-0.132, 0.183$]                  \\
                   &  Effect at AC$=1$ & $-0.006$;\ \        & CI[$-0.112, 0.102$]                  & \textbf{0.341};\ \ & \textbf{\textit{CI}[0.228, 0.471]} & $-0.026$;\ \        & CI[$-0.107, 0.052$]                  & \textbf{0.214};\ \ & \textbf{\textit{CI}[0.131, 0.309]} & 0.044;\ \                   &  CI[$-0.122, 0.210$]                  \\
                   \\ [-1ex]
    Time Spent     &  Index of MM    & 24.36;\ \          & CI[$-9.628, 58.76$]                  & $-5.954$;\ \         & CI[$-27.68, 15.16$]                  & 13.28;\ \          & CI[$-12.89, 42.17$]                  & 2.453;\ \          & CI[$-9.172, 14.36$]                  & \multicolumn{2}{c}{---} \\
                   &  Effect at AC$=0$ & 6.904;\ \          & CI[$-5.576, 21.55$]                  & 0.798;\ \          & CI[$-5.627, 7.986$]                  & 5.710;\ \          & CI[$-0.823, 16.28$]                  & $-0.723$;\ \         & CI[$-5.318, 2.348$]                  & 28.10;\ \                   &  CI[$-20.99, 77.18$]                  \\
                   &  Effect at AC$=1$ & \textbf{31.26};\ \ & \textbf{\textit{CI}[0.383, 62.91]} & $-5.157$;\ \         & CI[$-26.42, 15.00$]                  & 18.99;\ \          & CI[$-5.25, 46.89$]                   & 1.730;\ \          & CI[$-9.468, 12.84$]                  & \textbf{52.83};\ \          &  \textbf{\textit{CI}[3.981, 101.7]} \\
                   \\ [-1ex]
    Mot. Fut. Play &  Index of MM    & $-0.111$;\ \         & CI[$-0.250, 0.030$]                  & 0.096;\ \          & CI[$-0.025, 0.226$]                  & \textbf{0.174};\ \ & \textbf{\textit{CI}[0.068, 0.293]} & \textbf{0.388};\ \ & \textbf{\textit{CI}[0.157, 0.627]} & \multicolumn{2}{c}{---} \\
                   &  Effect at AC$=0$ & 0.039;\ \          & CI[$-0.011, 0.096$]                  & 0.051;\ \          & CI[$-0.006, 0.122$]                  & 0.006;\ \          & CI[$-0.024, 0.044$]                  & 0.096;\ \          & CI[$-0.055, 0.256$]                  & 0.034;\ \                   &  CI[$-0.155, 0.223$]                  \\
                   &  Effect at AC$=1$ & $-0.072$;\ \         & CI[$-0.202, 0.057$]                  & \textbf{0.147};\ \ & \textbf{\textit{CI}[0.046, 0.260]} & \textbf{0.180};\ \ & \textbf{\textit{CI}[0.081, 0.295]} & \textbf{0.484};\ \ & \textbf{\textit{CI}[0.313, 0.669]} & 0.070;\ \                   &  CI[$-0.128, 0.267$]                  \\
                   \\ [-1ex]
    Game Rec.      &  Index of MM    & $-0.020$;\ \         & CI[$-0.147, 0.111$]                  & 0.099;\ \          & CI[$-0.003, 0.209$]                  & 0.098;\ \          & CI[$-0.001, 0.209$]                  & \textbf{0.396};\ \ & \textbf{\textit{CI}[0.149, 0.645]} & \multicolumn{2}{c}{---} \\
                   &  Effect at AC$=0$ & 0.042;\ \          & CI[$-0.006, 0.097$]                  & 0.025;\ \          & CI[$-0.004, 0.071$]                  & 0.026;\ \          & CI[$-0.003, 0.073$]                  & 0.103;\ \          & CI[$-0.065, 0.275$]                  & $-0.014$;\ \                  &  CI[$-0.193, 0.165$]                  \\
                   &  Effect at AC$=1$ & 0.022;\ \          & CI[$-0.098, 0.143$]                  & \textbf{0.124};\ \ & \textbf{\textit{CI}[0.032, 0.229]} & \textbf{0.124};\ \ & \textbf{\textit{CI}[0.033, 0.229]} & \textbf{0.499};\ \ & \textbf{\textit{CI}[0.323, 0.685]} & 0.063;\ \                   &  CI[$-0.130, 0.255$]                  \\
    \bottomrule
    \end{tabular}}

    \vspace{1ex}
    \caption{Moderated mediation results for each path with the inclusion of the moderator $W$ (audial choice). For each variable and path, the index of moderated mediation (MM) and the conditional effects when Audial Choice (AC) is set to 0 and 1 are shown. Direct effects (\textsc{$X$$\rightarrow$$Y$}) do not have an index of moderated mediation. Significant results are bold. Significant results are based on 95\% \textit{CI}.}\label{tab:moderatedmediation}
\vspace*{-9pt}
\end{table*}

For the exploratory analyses with no a priori hypotheses, we test the model seen in Figure~\ref{fig:process59}. Results of the moderated mediation are found in Table~\ref{tab:moderatedmediation}. We find evidence of significant moderated mediation through the moderator of audial choice for intrinsic motivation (moderating \textsc{$X$$\rightarrow$$M_1$$\rightarrow$$Y$} and $X$$\rightarrow$$M_4$$\rightarrow$$Y$), immersion (moderating \textsc{$X$$\rightarrow$$M_2$$\rightarrow$$Y$} and $X$$\rightarrow$$M_4$$\rightarrow$$Y$), time spent playing (moderating \textsc{$X$$\rightarrow$$Y$}), motivation for future play (moderating $X$$\rightarrow$$M_3$$\rightarrow$$Y$ and $X$$\rightarrow$$M_4$$\rightarrow$$Y$), and likelihood of game recommendation (moderating $X$$\rightarrow$$M_4$$\rightarrow$$Y$). These effects were then probed while fixing the value of audial choice to 0 or 1 (see Table~\ref{tab:moderatedmediation}). When these effects were probed while fixing audial choice to 0, the mediations in \textit{all cases} were non-significant. On the other hand, when fixing audial choice to 1, the mediations in \textit{all cases} were positive and significant. Therefore, audial choice positively moderates different paths across all outcome variables.\footnote{It is worth noting that despite using a different model with the inclusion of the moderator $W$, there are overlaps with the results from hypothesis testing (Section~\ref{sec:hypothesistests}). For example, in all 7 cases of significant moderated mediation in this second model, indirect effects in the first model along those same paths are significant (see Table~\ref{tab:mediation}). Similarly, in all 5 cases where the index of moderated mediation was \textit{not} significant (or in the case of direct effects, non-existent), but the effect at AC=1 was significant, we have a significant effect in the first model along the same path. A slight divergence (2 cases) from the first model appears for the indirect effect $X$$\rightarrow$$M_3$$\rightarrow$$Y$. The indirect effect was significant in the first model for intrinsic motivation and time spent playing, but the effect is not significant at either value of AC (0 or 1) in the second model. Therefore, inclusion of the moderator $W$ does slightly affect the results from the model we chose for hypothesis testing. If we were to consider only the results from the second model, our overall hypothesis testing results would remain the same, but with slightly weaker support for H3.2 and H5.2.}\textsuperscript{,}\footnote{Another point to address is the possibility that the results arise because of participants who are randomly assigned a differently-gendered model. For example, perhaps audial choice alone is not effective at engendering outcomes specifically because gender is then randomly assigned, and will not match the player's gender \textasciitilde50\% of the time. To check if this was the case, we re-ran \textit{all} analyses with \textit{only} participants who self-identified as male and used a male model and participants who self-identified as female and used a female model regardless of condition (\textit{N}=808). We found identical results with respect to each of our hypotheses, and no evidence that random assignment of a differently-gendered avatar affected any of the results.\label{randomgenderaffectresults}}

\section{Discussion}

Existing work on avatar customization has focused almost exclusively on visual aspects of customization. While there are many benefits to avatar customization, it is unknown whether audial avatar customization confers similar benefits.

We conducted a 2 x 2 (visual choice x audial choice) experiment. Visual customization directly increases avatar identification and autonomy. Visual customization directly increases time spent playing and indirectly (through avatar identification and autonomy as mediators) increases intrinsic motivation,\footnote{An exception here is for the indirect effect of visual choice on intrinsic motivation through wishful identification, which was the \textit{only} significant result across all analyses with a negative effect. The reason for this is not immediately apparent, since the other indirect effects through wishful identification on time spent playing, motivation for future play, and likelihood of game recommendation were all significant and positive. One potential explanation is that when we identify wishfully with an avatar, we may view the avatar as a more competent version of ourselves (e.g., a better programmer). Even if we view the game as being interesting, this could detract from the enjoyability of the game (e.g., we feel ``less than'' our avatar), but not from intention to play or recommend. This is not the first time that such a discrepancy has been noted in the literature for wishful identification. Wishful identification has been found to be positively associated with PX outcomes but negatively associated with quality of created artifacts in an educational play and making context \cite{kaoEffects2018}. In other studies which measure wishful identification---e.g., in an  entertainment-oriented context  \cite{birkFostering2016}, no such negative associations have been found. It is possible that wishful identification (which is known to be correlated with lower psychological well-being \cite{Bessiere2007,higgins1987self,moretti1990relating}) has a more two-sided nature in educational contexts, potentially because they may feel more achievement-oriented rather than for ``fun.'' Additional controlled studies which manipulate game type and/or framing are needed to make more conclusive claims.} immersion, time spent playing, motivation for future play, and likelihood of game recommendation. Audial customization did not lead to a direct increase in avatar identification and autonomy. A significant interaction effect showed that audial customization directly increases avatar identification and autonomy, but only when visual customization was also available. Audial customization significantly moderated eight paths between visual customization and the outcome variables intrinsic motivation, immersion, time spent playing, motivation for future play, and likelihood of game recommendation. The moderation was such that when audial customization was unavailable, the path had a non-significant effect on the outcome, but when audial customization was available, the path had a significant effect on the outcome. Based on these results, we conclude that audial customization plays an important role in affecting outcomes.

However, we make the argument that although audial customization is important, it appears to have a weaker effect in comparison to visual customization. This argument is based on two facets of the results: (1) visual customization \textit{alone} has a significant effect on avatar identification and autonomy, whereas audial customization has a significant effect only within the group of participants who also have visual customization available;\footnote{Note that this is still the case even when only considering participants whose self-reported gender matched the avatar's as discussed in footnote \ref{randomgenderaffectresults}.} and (2) audial customization's effects on avatar identification and autonomy have lower effect sizes (small) when compared to visual customization (small-to-medium) \cite{cohen2013statistical,miles2001applying}. The first point suggests that audial customization plays an enhancing role for visual customization (i.e., when visual customization was present, audial customization further increased avatar identification and autonomy compared to no audial customization). Both points together suggest that audial customization, although important, is somewhat weaker than visual customization.

\subsection{Visual Customization Has a Stronger Effect Than Audial Customization}

Many possibilities exist for why visual customization had a stronger effect than audial customization. One possibility is that players are simply more familiar with visual customization. People are known to prefer things due to familiarity alone. The familiarity principle (also called the mere-exposure effect) describes the phenomenon of preference for things merely due to familiarity \cite{zajonc2001mere}. Therefore, the effects of visual customization could have been enhanced through familiarity.

Additionally, the total exposure time to the audial customization aspects of the avatar (i.e., voice) was only a fraction of the exposure time to the visual aspects of the avatar (i.e., model). While the audial aspect of the avatar is infrequent and typically only occurs before and after each puzzle, the visual aspect of the avatar is always present on screen. Moreover, the audial aspect of the avatar was interleaved with other sounds (game audio and background noise). Such factors could have all served to reduce the impactfulness of audial customization. Studying games with frequent voice lines (e.g., a narrative adventure such as \textit{The Walking Dead} \cite{WalkingDead}) would help to balance the exposure between visual and audial aspects of the avatar. Such studies would help to understand if the reason for the discrepancy between visual and audial customization effects stems from exposure.

\enlargethispage{-12pt}

Visual aspects of an avatar might also inherently (at a fundamental level) be more important than audial aspects. Humans have been shown to have better visual memory than auditory memory and that there appear to be fundamental differences between visual and auditory processing \cite{cohen2009auditory}. The \textit{picture superiority effect} describes the phenomenon whereby pictures and images are more often remembered compared to words \cite{childers1984conditions}. Reasons for why the picture superiority effect happens are still being debated. However, this fundamental asymmetry between visual and auditory stimuli would give credibility to the argument that visual aspects of an avatar are inherently more important than audial aspects of the avatar.

It may also be possible to explain the audial-visual discrepancy through investment of effort. If participants view the visual aspect of their avatar as more important, then they may invest more effort into visual customization than audial customization. According to Cialdini's commitment and consistency principle, people tend to behave in ways consistent with how they have acted in the past \cite{Cialdini1999} (i.e., future behavior often resembles past behavior). To maintain consistency with the effort in customizing the avatar visually, players would also invest more effort into the game. This would increase outcomes (e.g., avatar identification). Future studies could study the customization process itself more closely---e.g., time spent on customizing visual vs. audial aspects, measuring cognitive load in customizing visual vs. audial aspects.
\subsection{Audial Customization Is Effective When Paired With Visual Customization, But Not Alone}

Interestingly, audial customization was only effective at increasing avatar identification and autonomy when visual customization was also present. This was true even when we re-performed all analyses with only participants with a matching avatar gender (see footnote \ref{randomgenderaffectresults}). The reason for this is not immediately apparent. Although the character customization conditions were designed carefully and validated with expert UI designers, it is possible that the ability to customize voice (and especially in the absence of model selection) did not match players' expectations. A more in-depth investigation into the avatar customization process itself may help shed light on this phenomenon. Based on our results, we recommend pairing audial customization options with visual customization options to enhance outcomes.

\subsection{Implications for Research on Avatar Customization}

This research has examined both the effects of avatar customization (e.g., \cite{turkay2015effects,birkFostering2016}) and avatar customization interfaces (e.g.,  \cite{mcarthurAvatar2015,mcarthurMaking2019,mcarthurUX2017,paceAre2009}). Our contribution is a large-scale preregistered study showing that audial avatar customization, when paired with visual avatar customization, engenders important outcomes. Audial avatar customization was effective in increasing \textit{all} types of avatar identification (similarity, embodied, wishful) beyond the degree of avatar identification induced by visual avatar customization alone. Although prior studies (and the current study) show that \textit{visual} customization is effective at increasing avatar identification, we show that audial customization (in the form of a minimal set of voices) can also influence all aspects of avatar identification. This result suggests that even simple audial avatar customization—the selection of one voice from two options—is sufficient to increase perceived similarity with the avatar, the sense of being embodied within the avatar, and the idealization of the avatar. These three elements of identification are sometimes but not always consistently influenced by facets of avatar use \cite{vanlooyPlayer2012}, so this finding is particularly notable. Further, additional audial avatar customization options (e.g., pitch, loudness, pace, resonance, intonation) might facilitate even higher levels of avatar identification. Future studies could also investigate audial avatar customization in additional domains (e.g., exercise applications \cite{aloba2020toward}, social media and VR \cite{westerman2015effects,kolesnichenko2019understanding}), using additional methodological techniques (e.g., player interviews  \cite{banks2016avatars}), and the social inclusivity of audial customization interfaces (e.g., gender and race \cite{mcarthurAvatar2015,Wauck2018}).

The finding that audial choice significantly moderates the effect of visual choice on game outcomes (as mediated by identification and autonomy) provides further evidence for the importance of audial avatar customization. For example, visual customization was associated with greater intrinsic motivation (finding the game satisfying), sense of immersion, motivation for future play, and game recommendation, but only when there was audial customization, and all of these associations were fully mediated by embodied identification. In other words, visual customization alone did not sufficiently induce an association between embodied identification and these game outcomes, but visual together with audial customization did. Similarly, visual and audial customization together induced greater time spent playing the game and this effect was partially mediated by similarity identification. Together, these findings suggest that audial customization is a notable contributor not only to the subjective experience of identification with the avatar, but also to the \textit{outcomes} of identification with the avatar within the game. 

This work is also relevant to the Proteus Effect, a phenomenon whereby users tend to conform to the expected behaviors of their avatars \cite{Yee2007a}. This has been studied extensively with respect to visual characteristics \cite{ratanAvatar2020}, but not audial characteristics. For example, physically healthy-looking avatars can promote physical activity \cite{liWii2014}, and avatars perceived as creative can promote creative brainstorming \cite{gueganAvatarmediated2016}. However, allowing users to create audial avatar identities could also be a powerful avenue for inducing the Proteus Effect. In the present context of learning games, this research suggests that using an avatar with a voice that sounds more  capable of success in a computer-science context (e.g., intelligent, persistent) might empower players to perform better in the game and thus learn the educational content more effectively. Further research could be designed to confirm this expectation first by pretesting the perceived intelligence/persistence of different voices and then assigning exemplary voices as customization options within a similar game.

\subsection{Potential Applications for Audial Customization}

The amount of dialogue in CodeBreakers can be considered minimal compared to most games that contain voiced\break dialogue---e.g., Mass Effect \cite{MassEffect}. Nevertheless, audial avatar customization promoted all outcomes studied (e.g., autonomy, intrinsic motivation, immersion). Games for learning, health, and entertainment would all benefit from increases in the outcomes studied. Audial customization could have even broader implications in real-world devices. Examples include in-home devices incorporating voice interaction \cite{garg2020he,garg2020conversational,garg2021learn} and conversational agents more generally (e.g., Alexa \cite{amazon}) \cite{hiniker2021can,beneteau2020parenting,beneteau2019communication}. For example, it is not well understood what the effects of changing the voices of these devices are (e.g., to be more similar to the user). This includes other companion devices, such as robotic learning companions \cite{lubold2021effects,tian2020understanding,zuckerman2020companionship} and other dialogue-capable digital agents \cite{buddemeyer2021agentic,kim2019comparing,landoni2019my}.

Audial customization could enhance video instruction \cite{chang2019design,morrison2014khan} (e.g., lecture videos \cite{kizilcec2014showing,kizilcec2015instructor,monserrat2013notevideo}), massive open online courses (MOOCs) \cite{gamage2015quality}, intelligent tutoring \cite{chi2014tutorial,nelson2017comprehension}, e-books \cite{colombo2014design,rubegni2021girl}, and collaborative platforms \cite{kim2021winder,kumar2007supporting}. This could involve different modalities such as tangibles \cite{fan2018block}, tabletop displays \cite{kharrufa2010digital,maldonado2012interactive}, interactive installations \cite{long2018don,roberts2018digital,rubegni2020child}, augmented reality \cite{bonsignore2016traversing,cai2014case} and virtual reality \cite{lui2020immersive,aymerich2014relationship,freeman2021body}, and digital streaming \cite{pellicone2017game,chen2021afraid}. Additional investigation into different domains and modalities would elucidate whether audial customization can be applied more generally to increase user engagement. It is also important to investigate how the design choices behind audial customization can influence user identities---e.g., underrepresented minorities in STEM \cite{ahn2014want}---and how those design choices can be either exclusionary or inclusionary  \cite{kafai2017diversifying,richard2018gendered} and influence phenomena including stereotype threat \cite{richard2013investigating,ratan2015leveling} and user anxiety \cite{pimentelYour2020}. Further research is needed on audial customization to understand more generally the potential use cases.

\section{Limitations}

Despite the robust design of this controlled experiment, there are some limitations to the study's external and internal validity that should be considered in future research. First, participants were given only two visual and audial customization choices for each gender. Many games provide a greater number of choices during avatar customization, suggesting that avatar identification in such games is generally higher than it was in our study. Further, participants were likely more familiar with visual avatar customization than audial customization given that the former is more prevalent in current games and social media. Hence, the choice of avatar appearance—even based on just two options—was more likely to remind participants of previous avatar customization experiences that involved choices over many visual aspects of an avatar. In contrast, audial customization could potentially include a wide range of avatar characteristics that were not included in the present study (e.g., footsteps, whistling, grunting noises, pitch modification), but the participants' choice of avatar voice was less likely to remind them of these possibilities. Moreover, avatar identification may have been limited for players who do not conform to stereotypical representations of ``male'' and ``female'' voices. For these reasons, future research on this topic should include a larger set of customization options, especially for audial avatar characteristics.

The study also included a potential confound relating to the attention paid to audial and visual cues. Namely, in order to proceed in the game, participants were required to solve visual puzzles that did not include audial elements. This prioritization of visual stimuli may have led to a greater focus on the avatar's appearance compared to avatar's speech, partially explaining why visual avatar customization was more consequential in the study outcomes. Another related but minor issue is that the quality of the sound hardware may have varied between players' computers causing noise in the data (i.e., less attention to audial cues), but this was likely not confounded with experiment condition given random assignment. Further, all participants performed an audio check, so a threshold of audial attention can be inferred.

The study relied on participants being paid to play the game, like most research in this field, which potentially limits ecological validity. Further, generalizability was not established beyond the single, education-oriented game designed for this research. Relatedly, the study cannot determine how specific facets of this particular game design (e.g., pacing) influenced the study outcomes. For one, the game was designed to highlight the avatar's voice for a single user, so the study findings do not directly speak to multi-user games which offer voice-based communication \cite{wadley2015voice,wadley2007speaking}. However, algorithmic voice modification (e.g., pitch modulation to mask gender) is an increasingly popular multi-user technology for games (e.g., \cite{Mayor2009,Voicemod2020}) that could potentially help mitigate toxic behavior between players \cite{Vella2020,Wadley2009}. The present findings indirectly suggest that customizing such voice modification might also be beneficial to the user's experience in other ways.

The study required participants to play the game for a minimum of 10 minutes, which is significantly less time than many people tend to play video games \cite{ESA2021}. However, this length of exposure is sufficient to induce avatar identification \cite{downsPolythetic2019}, as other studies have found \cite{Allen2021}, and the present study was not intended to examine changes in identification over time. We should also note that 10-minute exposures are common in video-game experiments, perhaps due to operational constraints, but these studies tend to find sufficient effects on their outcomes of interest with such durations.

\section{Conclusion}
Avatar customization is known to positively affect crucial outcomes in numerous domains, including health, entertainment, and education. However, studies on avatar customization have focused almost exclusively on  visual aspects of customization. It is unknown whether audial customization can confer the same benefits as visual customization. We presented one of the first studies to date on audial avatar customization. Participants with visual choice experienced higher avatar identification and autonomy. Participants with audial choice experienced higher avatar identification and autonomy, but only within the group of participants who had visual choice available. Visual choice led to an increase in time spent and indirectly led to  increases in intrinsic motivation, immersion, time spent, future play motivation, and likelihood of game recommendation. Audial choice moderated the majority of these effects. Our results suggest that audial customization, although having a moderately weaker effect compared to visual customization, plays an important role in enhancing all outcomes compared to visual customization alone. We discussed the implications for research and potential applications of audial avatar customization. This work takes an important first step in developing a baseline understanding of audial avatar customization.

\balance
\bibliographystyle{ACM-Reference-Format}
\bibliography{sample-base}

\appendix

\end{document}